\documentclass[smallcondensed]{svjour3}     
\smartqed  

\RequirePackage{fix-cm}

\AtBeginDocument{%
  \paperwidth=\dimexpr
    1in + \oddsidemargin
    + \textwidth
    + 1in + \oddsidemargin
  \relax
  \paperheight=\dimexpr
    1in + \topmargin
    + \headheight + \headsep
    + \textheight
    + 1in + \topmargin
  \relax
  \usepackage[pass]{geometry}\relax
}
\usepackage{cite}
\usepackage{amsmath,amssymb,amsfonts}
\usepackage{algorithmic}
\usepackage{graphicx}
\usepackage{textcomp}
\usepackage{soul}	
\usepackage{balance}
\usepackage{tabularx}
\usepackage{enumitem}
\usepackage{adjustbox}
\usepackage{booktabs}
\usepackage{multirow}
\usepackage{caption}
\usepackage{hyperref}
\usepackage{dirtytalk} 
\usepackage{xcolor}
\usepackage[most]{tcolorbox}
\usepackage{pifont}
\usepackage[mathlines,switch]{lineno}
\newcommand*\circled[1]{\tikz[baseline=(char.base)]{%
            \node[shape=circle,draw,inner sep=1pt] (char) {#1};}}
\usepackage{tikz}
\usepackage{graphicx}
\usepackage{listings}
\usepackage{pifont}
\usepackage{amsmath}
\usepackage{pdflscape} 
\usepackage{tikz}
\usepackage{dirtytalk}
\usepackage{makecell}
\usepackage[title]{appendix}
\usepackage{pythonhighlight}
\usepackage{caption}

\definecolor{LightGray}{gray}{0.9}
\usepackage[autostyle]{csquotes}
\usepackage{longtable}
\usepackage{booktabs}
\usepackage{tablefootnote}
\usepackage{multirow}
\usepackage{adjustbox}
\usepackage{enumitem}
\usepackage{siunitx}
\usepackage{fancyvrb}
\usepackage{soul}
\usepackage{textcomp}
\usepackage{tabularx}
\newlength\someheight
\usepackage{pgfplots}
\usepackage{pgfplotstable}
\pgfplotsset{compat=1.14}
\definecolor{storeClusterComponent}{HTML}{808080}
\definecolor{dbscan}{HTML}{BEBEBE}
\definecolor{constructCluster}{HTML}{DCDCDC}
\usepackage{tikz} 
\usepackage{subcaption}
\usetikzlibrary{shapes,calc,shadows,arrows}
\usepackage{pgf-pie}
\usepackage{etoolbox}
\newtoggle{showpct}
\usepackage{xcolor,colortbl}
\usepackage[english]{babel} 

\definecolor{codegreen}{rgb}{0,0.6,0}
\definecolor{codegray}{rgb}{0.5,0.5,0.5}
\definecolor{codepurple}{rgb}{0.58,0,0.82}
\definecolor{backcolour}{rgb}{0.95,0.95,0.92}

\usepackage[most]{tcolorbox}

\newtcbtheorem{Summary}{\bfseries Summary}{enhanced,drop shadow={black!50!white},
  coltitle=black,
  top=0.3in,
  attach boxed title to top left=
  {xshift=1.5em,yshift=-\tcboxedtitleheight/2},
  boxed title style={size=small,colback=pink}
}{summary}

\newtcolorbox[auto counter]{summary}[1][]{title={\bfseries Summary~\thetcbcounter},enhanced,drop shadow={black!50!white},
  coltitle=black,
  top=0.3in,
  attach boxed title to top left=
  {xshift=1.5em,yshift=-\tcboxedtitleheight/2},
  boxed title style={size=small,colback=pink},#1}

\usepackage[round]{natbib}

\usepackage{etoolbox}
\makeatletter
\patchcmd{\pgfpie@slice}%
{\scalefont{#3}\beforenumber#3\afternumber}%
{\iftoggle{showpct}{\scalefont{#3}\beforenumber#3\afternumber}{}}%
{}{}

\patchcmd{\NAT@citex}
  {\@citea\NAT@hyper@{%
     \NAT@nmfmt{\NAT@nm}%
     \hyper@natlinkbreak{\NAT@aysep\NAT@spacechar}{\@citeb\@extra@b@citeb}%
     \NAT@date}}
  {\@citea\NAT@nmfmt{\NAT@nm}%
   \NAT@aysep\NAT@spacechar\NAT@hyper@{\NAT@date}}{}{}

\patchcmd{\NAT@citex}
  {\@citea\NAT@hyper@{%
     \NAT@nmfmt{\NAT@nm}%
     \hyper@natlinkbreak{\NAT@spacechar\NAT@@open\if*#1*\else#1\NAT@spacechar\fi}%
       {\@citeb\@extra@b@citeb}%
     \NAT@date}}
  {\@citea\NAT@nmfmt{\NAT@nm}%
   \NAT@spacechar\NAT@@open\if*#1*\else#1\NAT@spacechar\fi\NAT@hyper@{\NAT@date}}
  {}{}

\def\srcfile[#1,#2]#3{
    \node [draw, fill=white, minimum height=1.7cm, minimum width=1.35cm, rounded corners, double copy shadow={shadow xshift=0.1cm, shadow yshift=0.1cm}] (#1) at #2 {};
    \node[align=center, font=\sffamily\fontsize{5}{1.5}\selectfont] at #2 {#3};
}

\makeatother

\definecolor{dkgreen}{rgb}{0,0.6,0}
\definecolor{gray}{rgb}{0.5,0.5,0.5}
\definecolor{mauve}{rgb}{0.58,0,0.82}
\definecolor{dgreen}{rgb}{0.0, 0.5, 0.0}

\usepackage{mdframed}
\newcounter{finding}
\newmdenv[%
    linewidth=0.6pt,
    linecolor=black,
    outerlinewidth=0pt,
    skipabove=0pt,
    skipbelow=0pt,
    settings={\global\refstepcounter{finding}},
]{myfinding}

\makeatletter
\newcommand\notsotiny{\@setfontsize\notsotiny\@vipt\@viipt}
\makeatother

\usepackage{amsthm}

\theoremstyle{definition}

\usepackage{threeparttable}

\usepackage{array}
\definecolor{dodgerblue}{RGB}{30,144,255}
\definecolor{orange}{RGB}{255, 120, 8}

\newcommand{\lstbg}[3][0pt]{{\fboxsep#1\colorbox{#2}{\strut #3}}}
\lstdefinelanguage{diff}{
  basicstyle=\ttfamily\small,
  morecomment=[f][\lstbg{red!20}]-,
  morecomment=[f][\lstbg{green!20}]+,
  morecomment=[f][\textit]{@@},
}

\newcolumntype{C}[1]{>{\centering\arraybackslash}m{#1}}

\tcbset{commonstyle/.style={boxrule=0pt,sharp corners,enhanced jigsaw,nobeforeafter,boxsep=0pt,left=\fboxsep,right=\fboxsep}}
\newtcolorbox{mycolorbox}[1][]{commonstyle,#1}

\usepackage{svg}

\usepackage{etoolbox}
\newcommand*{\affaddr}[1]{#1} 
\newcommand*{\affmark}[1][*]{\textsuperscript{#1}}

\usepackage{scalerel}
\usepackage{tikz}
\usetikzlibrary{svg.path}

\definecolor{orcidlogocol}{HTML}{A6CE39}
\tikzset{
  orcidlogo/.pic={
    \fill[orcidlogocol] svg{M256,128c0,70.7-57.3,128-128,128C57.3,256,0,198.7,0,128C0,57.3,57.3,0,128,0C198.7,0,256,57.3,256,128z};
    \fill[white] svg{M86.3,186.2H70.9V79.1h15.4v48.4V186.2z}
                 svg{M108.9,79.1h41.6c39.6,0,57,28.3,57,53.6c0,27.5-21.5,53.6-56.8,53.6h-41.8V79.1z M124.3,172.4h24.5c34.9,0,42.9-26.5,42.9-39.7c0-21.5-13.7-39.7-43.7-39.7h-23.7V172.4z}
                 svg{M88.7,56.8c0,5.5-4.5,10.1-10.1,10.1c-5.6,0-10.1-4.6-10.1-10.1c0-5.6,4.5-10.1,10.1-10.1C84.2,46.7,88.7,51.3,88.7,56.8z};
  }
}

\newcommand\orcidicon[1]{\href{https://orcid.org/#1}{\mbox{\scalerel*{
\begin{tikzpicture}[yscale=-1,transform shape]
\pic{orcidlogo};
\end{tikzpicture}
}{|}}}}

\usepackage{listings}
\usepackage{xcolor} 

\lstdefinestyle{python}{
    language=Python,
    basicstyle=\ttfamily\tiny,
    keywordstyle=\color{blue}\bfseries,
    commentstyle=\color{green!50!black}\itshape,
    stringstyle=\color{red},
    showstringspaces=false,
    frame=single,
    numbers=left,
    numberstyle=\tiny\color{gray},
    stepnumber=1,
    numbersep=5pt,
    breaklines=true,
    breakatwhitespace=true,
    tabsize=4,
    captionpos=b
}

\usepackage{hyperref}

\journalname{Empirical Software Engineering}
\begin{document}

\title{Logging Requirement for Continuous Auditing of Responsible Machine Learning-based Applications}

\author{Patrick Loic Foalem \and Leuson Da Silva \and Foutse Khomh \and Heng Li \and Ettore Merlo
}

\authorrunning{Patrick Loic Foalem\and Leuson Da Silva \and Foutse Khomh \and Heng Li \and Ettore Merlo}


\institute{ \affmark[*]Corresponding author. \\
\\
           Patrick Loic Foalem \and  Leuson Da Silva \and Foutse Khomh \and Heng Li \and  Ettore Merlo \at
              \affaddr{Department of Computer Engineering and Software Engineering \\ Polytechnique Montreal \\
              Montreal, QC, Canada} \\
              \email{\{patrick-loic.foalem \and  leuson-mario-pedro.da-silva \and  foutse.khomh \and  heng.li \and ettore.merlo\}@polymtl.ca}           
}

\date{Received: date / Accepted: date}

\maketitle

\section*{Abstract}
Machine learning (ML) is increasingly used across various industries to automate decision-making processes. However, concerns about the ethical and legal compliance of ML models have arisen due to their lack of transparency, fairness, and accountability. Monitoring, particularly through logging, is a widely used technique in traditional software systems that could be leveraged to assist in auditing ML-based applications. Logs provide a record of an application's behavior, which can be used for continuous auditing, debugging, and analyzing both the behavior and performance of the application.

In this study, we investigate the logging practices of ML practitioners to capture responsible ML-related information in ML applications. We analyzed 85 ML projects hosted on GitHub, leveraging 20 responsible ML libraries that span principles such as privacy, transparency \& explainability, fairness, and security \& safety. Our analysis revealed important differences in the implementation of responsible AI principles. For example, out of 5,733 function calls analyzed, privacy accounted for 89.3\% (5,120 calls), while fairness represented only 2.1\% (118 calls), highlighting the uneven emphasis on these principles across projects. Furthermore, our manual analysis of 44,877 issue discussions revealed that only 8.1\% of the sampled issues addressed responsible AI principles, with transparency \& explainability being the most frequently discussed principles (32.2\% of all issues related to responsible AI principles).

Additionally, a survey conducted with ML practitioners provided direct insights into their perspectives, informing our exploration of ways to enhance logging practices for more effective, responsible ML auditing. We discovered that while privacy, model interpretability \& explainability, fairness, and security \& safety are commonly considered, there is a gap in how metrics associated with these principles are logged. Specifically, crucial fairness metrics like group and individual fairness, privacy metrics such as epsilon and delta, and explainability metrics like SHAP values are not considered current logging practices.

The insights from this study highlight the need for ML practitioners and logging tool developers to adopt enhanced logging strategies that incorporate a broader range of responsible AI metrics. This adjustment will facilitate the development of auditable and ethically responsible ML applications, ensuring they meet emerging regulatory and societal expectations. These specific insights offer actionable guidance for improving the accountability and trustworthiness of ML systems.


\section{Introduction}
\label{sec:introduction}

Machine learning-based applications are increasingly adopted across various industries, including healthcare, finance, hiring, and criminal justice, to automate decision-making processes \citep{mehrabi2021survey}. While these applications offer efficiency and scalability, they also raise concerns regarding transparency, fairness, and accountability. Responsible AI aims to address these concerns by ensuring AI systems adhere to societal, ethical, and legal standards. Consequently, there is a growing demand for auditing and regulating machine learning applications to ensure compliance with ethical and legal requirements \citep{AIActEU2023}.

An essential technique for auditing machine learning-based applications is monitoring \citep{anderson1993auditing}, particularly in continuous auditing settings \citep{hazar2021new, alles2006continuous, singh2014continuous}. Logging is a widely used approach for monitoring both traditional \citep{chen2020studying, liu2009framework, takada2002tudumi} and machine learning applications \citep{foalem2024studying}. Logging involves capturing and storing relevant information about the application's inputs, outputs, and internal decisions. These logs serve as a record of the system’s behavior, facilitating auditing, debugging, and performance analysis \citep{zeng2019studying, chen2019empirical}.

Various tools and techniques have been developed to audit machine learning-based applications and ensure their alignment with responsible AI principles \citep{hopkins2021machine, jagielski2020auditing, dickey2019machine}. One key practice commonly used for such audits is logging, which helps collect information about the behavior of AI systems \citep{kuna2014outlier}. Weights \& Biases (W\&B), for instance, is a popular machine learning experiment tracking tool that enables practitioners to log performance metrics for their models, ensuring better monitoring and reproducibility.
For example, W\&B is often used to log the Root Mean Squared Error (RMSE), a key evaluation metric that measures the square root of the mean squared differences between predicted and actual values. Below is an example of logging RMSE values for both training and validation phases, including out-of-distribution (OOD) performance:
\\

\begin{lstlisting}[language=Python, caption=Example of logging metrics using Weights \& Biases.]
if USE_WANDB:
    wandb.log({"RMSE_val": np.sqrt(np.mean(val_losses)), 
               "RMSE_training": np.sqrt(np.mean(losses))}, 
               commit=(not args.val_ood))
if USE_WANDB:
    wandb.log({"RMSE_valOOD1": np.sqrt(np.mean(val_losses1))})
print("Root Mean Squared Validation OOD1 Error =", np.sqrt(np.mean(val_losses1)))
if USE_WANDB:
    wandb.log({"RMSE_valOOD3": np.sqrt(np.mean(val_losses3))})
if USE_WANDB:
    wandb.log({"RMSE_valOOD4": np.sqrt(np.mean(val_losses4))}, commit=True)

\end{lstlisting}

This example illustrates how logging is used in ML projects \footnote{\url{https://shorturl.at/QVUzf}} to track performance metrics such as RMSE. However, while logging for performance monitoring is common, our study reveals a critical gap--current logging practices fail to capture essential responsible AI metrics such as fairness, explainability, and privacy. This shortcoming limits the ability of practitioners to audit ML systems for compliance with responsible AI principles effectively.

In contrast, traditional software systems rely on audit logs \footnote{\url{https://www.datadoghq.com/knowledge-center/audit-logging}} to document compliance and system behavior using logging and tracing techniques \citep{dumais2014understanding}. These logs are frequently used for policy enforcement, security monitoring, and failure analysis \citep{lee2013loggc, soderstrom2013secure}. While traditional logging frameworks provide valuable insights, logging in machine learning presents unique challenges that require specialized approaches to support responsible AI auditing.


To address this gap, our study investigates the use of logging practices for collecting responsible AI-related information in ML applications. Understanding how ML practitioners record this information is crucial, as it can be used for auditing or provide insights for audit requirements. Our focus is on the information obtained using responsible AI libraries, which support measuring and correcting issues impacting ML applications' safety, privacy, transparency, interpretability, fairness, and accountability.

This study aims to provide insights into the current practices of ML practitioners regarding the use of logging to capture system behavior information pertinent to responsible AI. It assesses the adequacy of existing logging mechanisms in supporting auditing activities and facilitates the development of recommendations to enhance logging practices in ML applications for auditing purposes. We seek answers to the following research questions:
\begin{itemize}
    \item[] \textit{RQ1: What responsible AI principles are considered by ML practitioners?}
    \item[] \textit{RQ2: What runtime information of the ML applications is logged to support responsible AI auditing?}
    \item[]  \textit{RQ3: What specific logging requirements and information are essential for effective auditing of responsible AI principles?} 
\end{itemize}

We analyzed 85 ML projects from GitHub that show evidence of the consideration of responsible AI principles in their source code or issue discussions. Within these projects, we investigated 5,733 function calls from 20 responsible AI libraries and found significant disparities in the implementation of responsible AI principles. Privacy accounted for 89.3\% of the function calls (5,120 calls), while fairness accounted for only 2.1\% (118 calls), highlighting an uneven emphasis on these principles across projects. Additionally, our manual analysis of 44,877 reported issues revealed that only 8.1\% of the sampled issues (31 of 382 issues) addressed responsible AI principles, with transparency \& explainability emerging as the most frequently discussed principles (32.2\% of all responsible AI issues).

We further analyzed logged information in these responsible AI projects to identify gaps between logged information and responsible AI requirements. Through this analysis, we identified a set of responsible AI metrics, such as fairness metrics (e.g., group and individual fairness), privacy metrics (e.g., epsilon and delta), and explainability metrics (e.g., SHAP values), that should be under continuous surveillance and documentation within ML applications. Recognizing the need for a structured approach to guide this process, we developed a taxonomy of logging practices relevant to auditing responsible AI principles in ML applications. This taxonomy, presented as a logging diagram for responsible ML applications, was further refined and validated through a survey with ML practitioners, which yielded a participation rate of 0.8\% (22 responses). Practitioners feedback directly confirmed the relevance of the identified metrics and the practicality of the proposed logging practices to develop responsible ML applications.

{To the best of our knowledge, this is the first study examining the role of logging in ML auditing and providing a comprehensive framework for addressing the gaps in logging responsible AI principles.

The primary contributions of this work are as follows:
\begin{enumerate}
    \item We provide empirical evidence that privacy, model interpretability \& explainability, fairness, and security \& safety are the most common aspects considered by practitioners when implementing responsible AI principles in their projects.
    \item Our results show that ML practitioners do not log metrics related to responsible AI principles.
    \item We propose a logging framework for responsible ML-based applications that outline metrics that should be continuously monitored to ensure adherence to responsible AI principles. This framework was enriched by contributions from our survey, where practitioners shared their insights on the relevance and implementation of our proposed framework.
    \item We make our replication package available \footnote{\url{https://github.com/foalem/audit-responsible-ml-paper}}.
\end{enumerate}
Our observations offer valuable insights for ML practitioners seeking to improve their logging practices by monitoring responsible AI metrics. Additionally, our study emphasizes the need to log beyond conventional information, such as model performance, hyperparameters, and data preprocessing, as outlined by \citet{foalem2024studying}, to build auditable and responsible ML applications.
Our findings can also guide logging tool developers in enhancing their tools or creating new ones to support the logging of responsible AI principles and facilitate the auditing of ML applications through logging.\\
\textbf{Organization:} The paper is structured as follows: Section \ref{sec:background} presents background information about machine learning audit, different audit techniques, and Responsible AI principles, while Section \ref{sec:approach} explains the experimental setup. The research question results are presented in Section \ref{sec:result}, followed by a discussion of these results in Section \ref{sec: discussion_implication}. Section \ref{sec:related_work} provides a summary of related works. Section \ref{sec:threats_to_validity} discusses potential threats to the validity of the study. Finally, in Section \ref{sec:conclusion}, we present our conclusions and suggest avenues for future research.

\section{Background}
\label{sec:background}
This section introduces key concepts essential for understanding the work outlined in the paper.

\subsection{Audit in machine learning}
Audit in machine learning involves a systematic examination and evaluation of the performance and behavior of machine learning algorithms and models, to ensure that they are operating as intended, without bias or unintended consequences \citep{Censius2023AIAudit}. This evaluation is typically performed to examine the inputs, outputs, and decision-making process of the algorithm or model, as well as the quality and appropriateness of the data used to train the model. The purpose of the audit is to ensure that the machine learning system is reliable, accurate, fair, and transparent \citep{clark2018machine}. The audit process may involve various techniques and tools, including monitoring, testing, and validation, and it may cover various aspects of the machine learning system, such as data privacy, model interpretability, fairness, and security \citep{clark2018machine, yang2024robustness}. The audit results may be used to identify and mitigate any issues or risks associated with the machine learning system, as well as to improve its performance and usability. Based on the above, the key motivation for our research is to investigate how practitioners monitor responsible AI-related metrics that could be leveraged for auditing compliance with responsible AI principles.

\subsection{Auditing techniques}
In the area of information technology (IT), the audit is the process of evaluating the effectiveness, efficiency, and security of an organization's information systems \citep{otero2018information}. IT audit plays a crucial role in ensuring that organizations' information systems are operating effectively, efficiently, and securely to achieve the desired goals.
To assess the effectiveness of an organization's information systems, IT auditors use various techniques and tools. Some commonly used techniques include:
\begin{itemize}[leftmargin=0.8 pt,align=left]
    \item \textbf{Risk-based auditing}: This technique involves identifying the most significant risks to an organization's information systems and directing the audit effort towards those areas. By prioritizing their work, auditors ensure that they focus on the most critical aspects of the system.\citep{griffiths2012risk}.
    \item \textbf{Compliance auditing}: This technique involves assessing the organization's compliance with laws, regulations, and industry standards. Compliance auditing is essential to ensure that organizations meet legal, regulatory requirements, and certification standards \citep{thottoli2021relevance}. 
    \item \textbf{Process-based auditing}: This technique involves evaluating the organization's processes to ensure they are efficient, effective, and meet business objectives. Process-based auditing helps auditors to identify areas for improvement and recommend process changes to improve the overall effectiveness of the system \citep{guldenmund2006development}.
\end{itemize}
ML auditing requires a tailored set of techniques and tools since it has its unique challenges and more specific auditing requirements, such as the need for explainability and transparency of the decision-making process and the need for monitoring the data and algorithmic bias.

\subsection{Continuous auditing}
Continuous auditing (CA) and continuous monitoring (CM) represent integral components of a robust framework for ensuring responsible AI practices. Continuous auditing, as detailed in the Global Technology Audit Guide (GTAG), is an ongoing process of risk and control assessments facilitated by advanced technology. It transitions from periodic to continuous evaluations, encompassing a wide spectrum of transactions and data sources, such as security levels, logging incidents, and changes in IT configurations \citep{gtag}. This methodology allows internal audit functions to offer real-time assurance on effective risk management and control frameworks, substantially enhancing efficiency and providing deeper insights.

Continuous monitoring, stands as a cornerstone for management, enabling an uninterrupted assessment of control effectiveness and risk identification within business processes. It aims to refine business activities in alignment with ethical and compliance standards, fostering more informed risk-related decisions. When harmonized, CA/CM bolsters an organization's capability to identify and address issues in real-time, thus promoting a culture of transparency and accountability essential in the context of responsible AI applications \citep{gtag}.

Integrating continuous auditing into the development and oversight of AI applications, especially in logging metrics related to responsible AI principles like fairness, transparency, and accountability, is crucial. By continuously auditing AI systems for adherence to these principles, organizations can detect and mitigate ethical and operational risks more effectively \citep{deloitecacm}. This approach not only aligns with the increasing regulatory and societal demands for ethical AI but also supports the sustainable integration of AI technologies into business processes \citep{deloitecacm}. Moreover, it facilitates the provision of continuous assurance to stakeholders that AI applications operate within established ethical frameworks, thereby fostering trust and ensuring the long-term viability of AI initiatives.

\subsection{Responsible AI/ML principles}
\label{sec:resp}
Responsible AI/ML refers to the ethical and responsible development, deployment, and use of artificial intelligence and machine learning technologies \citep{deshpande2022responsible}. The growing adoption of AI/ML in various domains has raised concerns about their impact on society, including issues related to fairness, privacy, security, transparency, accountability, and explainability \citep{de2021companies}. Responsible AI/ML seeks to address these concerns and ensure that AI/ML technologies are developed and used in a way that is safe, ethical, and trustworthy. 
Responsible AI/ML involves several key principles, including fairness, accountability, transparency, privacy, and security \citep{de2021companies}.
\begin{itemize}
    \item \textbf{Fairness}: requires that AI/ML applications do not discriminate against any individual or group based on race, gender, age, or other factors.
    \item \textbf{Accountability}: refers to the responsibility of developers and users of AI/ML applications to ensure that they are used responsibly and ethically. 
    \item \textbf{Transparency \& Explainability}: requires that AI/ML applications be understandable and interpretable and that their decision-making processes can be explained to stakeholders.
    \item \textbf{Interpretability}: emphasizes the ability to understand and articulate how an AI/ML model generates its predictions or outputs, ensuring that its operations can be traced and justified.
    \item \textbf{Privacy}: requires that AI/ML applications protect the confidentiality of personal data and do not violate individuals' privacy rights.
    \item \textbf{Security \& Safety}: requires that AI/ML applications are secure and resistant to attacks or malicious use. 
\end{itemize}

Many organizations are now advocating for responsible AI; this is the case of the European Union (EU) which published a General Data Protection Regulation (GDPR) in 2018 to regulate the processing of personal data within the EU. This regulation includes provisions that require organizations to ensure that automated decision-making processes, including those based on AI, are transparent, explainable, and fair \citep{voigt2017eu, AIActEU2023}. The IEEE Global Initiative on Ethics of Autonomous and Intelligent Systems and the European Commission's AI have developed a set of guidelines to support the ethical design and deployment of autonomous and intelligent systems\citep{chatila2019ieee, neuwirth2022eu}. These guidelines include principles such as transparency, accountability, and privacy, and aim to promote the responsible use of AI \citep{chatila2019ieee, AIActEU2023}. In 2018, the Montreal Declaration for Responsible AI was also published to provide guidelines for the responsible development and deployment of AI. The Montreal Declaration includes principles such as transparency, accountability, and the protection of privacy and human dignity \citep{MontrealDeclaration2023}. 

However, implementing and ensuring adherence to these principles presents significant challenges. Without effective auditing mechanisms, operationalizing and enforcing these principles can be difficult. Our research aims to bridge this gap by investigating how metrics aligned with responsible AI principles are documented and utilized. Through this work, we aspire to contribute to the development of AI/ML technologies that not only advance in capability but also in their contribution to a just and ethical digital society.

\subsection{Responsible AI/ML library}
\label{sec:respon}
Responsible AI libraries serve as comprehensive frameworks or sets of functionalities that allow developers, researchers, and data scientists to assess, improve, and guarantee the ethical deployment of AI technologies. These libraries are grounded in the recognition that ensuring the responsible use of AI transcends mere compliance with existing legal frameworks; it involves a proactive engagement with the ethical dimensions of technology development.

The landscape of these tools is diverse, reflecting the multifaceted challenges of operationalizing responsible AI principles. From AIF360 to SHAP and OPACUS, each library offers unique approaches to diagnose, mitigate, and elucidate biases across different stages of ML model development from data collection and preprocessing to model training and evaluation. This variety not only accommodates the specific needs and contexts of different AI projects but also enriches the toolkit available to practitioners aiming to uphold ethical standards in their work.

These libraries offer feature model-agnostic capabilities, underscoring the universal applicability of responsible AI practices across various model architectures and applications. By providing metrics for fairness analysis, explanations for model decisions, and mechanisms for data privacy and security, these tools embody a practical response to the global call for ethical consideration in AI projects. 

In summary, responsible AI libraries are instrumental in bridging the gap between abstract ethical principles and concrete technical practices. They offer the means to operationalize responsible AI principles effectively, ensuring that AI and ML technologies are developed and deployed in ways that are not only innovative and efficient but also ethical and equitable. As such, these libraries play a pivotal role in the ongoing journey towards responsible AI, embodying the collective effort of the tech community to foster AI systems that are trustworthy, reliable, and aligned with human values.

\section{Experiment setup} 
\label{sec:approach}
In order to address the research questions investigated in this study, we adopt a mixed-method based on seven main steps, considering a combination of quantitative and qualitative approaches \citep{ivankova2006using}.  
Figure \ref{fig:researchprocess} provides an overview of our approach. In the following, we explain each step in detail, highlighting the techniques and methods used.
\begin{figure}[htp!]
    \centering
    \includegraphics[width=0.7\textwidth]{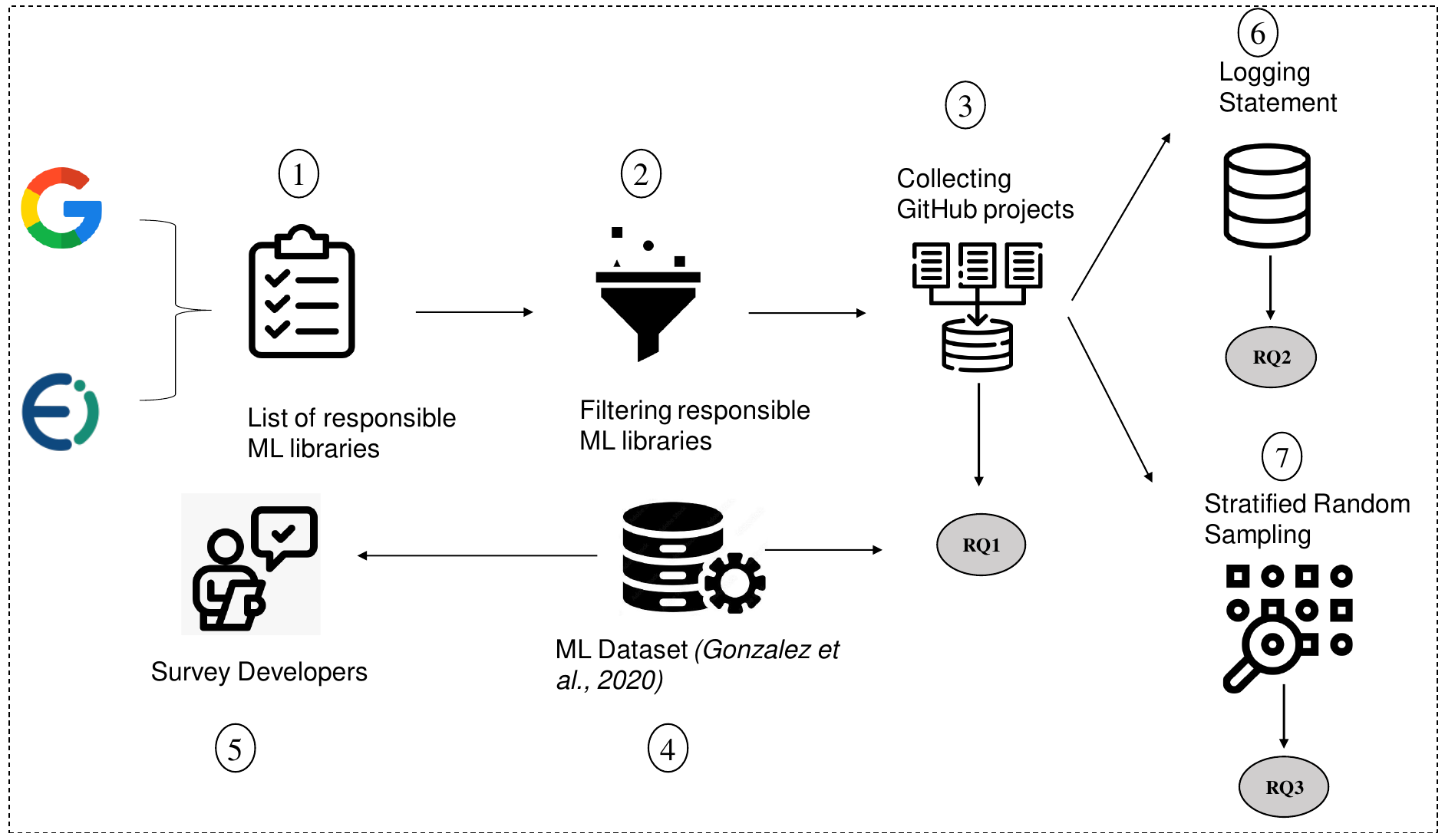}
    \caption{Overview of our research study.}
    \label{fig:researchprocess}
\end{figure}
\begin{enumerate}[label=\protect\circled{\arabic*},leftmargin=0.8 pt,align=left]
\item \textbf{Identifying a list of responsible ML libraries}: The purpose of this step is to comprehensively list responsible ML libraries recognized in both academic research and industry practice. We aim to obtain a 
holistic view of the landscape of responsible ML libraries. To identify responsible ML libraries, we performed a double search looking for research papers and industry tools. This approach allows us to cover a wide range of responsible ML libraries proposed in academia and the industry. The following describes the steps adopted to identify responsible ML libraries from each data source. 

\begin{itemize}
    \item \emph{Responsible ML libraries from research papers}:
\end{itemize}
To identify responsible ML libraries proposed in research papers, we conducted a systematic search using the following query (Responsible OR Fairness OR Privacy OR Auditing OR Transparency OR Security OR Explainability OR Interpretability) AND (Machine Learning OR ML) on the Google Scholar\footnote{\url{https://scholar.google.com/}} and Engineering Village\footnote{\url{https://www.engineeringvillage.com/}} platforms. Google Scholar was chosen for its broad coverage of multidisciplinary academic literature, while Engineering Village was selected for its specialized focus on engineering and technical research, ensuring comprehensive and domain-relevant results \citep{foalem2024studying}. The selection of these search terms was the result of collaborative discussions among all authors during a series of dedicated meetings \citep{tambon2022certify}. The terms were chosen to ensure a comprehensive focus on machine learning and explicitly align with the responsible AI properties described in Section \ref{sec:resp}, extracted from \citep{de2021companies}'s paper. These properties capture essential aspects such as fairness, privacy, transparency, explainability, and interpretability, ensuring that the search query encompasses a broad and relevant scope for identifying responsible ML libraries.
We made specific configurations on the research platforms to ensure relevant results, such as limiting scholarly literature published between 2010 and 2024 and setting up the Engineering Village to query Inspect and Compendex datasets.
For each source, we reviewed the first 20 returned research papers, as the results are sorted based on their relevancy \citep{foalem2024studying}. Since we wanted to perform an in-depth analysis, our focus was on responsible ML libraries hosted on GitHub to focus on open-source tools. Through our search process, we identified 12 open-source responsible ML libraries. 

\begin{itemize}
    \item \emph{Responsible ML libraries from Google search:}
\end{itemize} 
To identify open-source responsible ML libraries mentioned across various websites, we conducted a Google search using the query: (Machine Learning OR ML) AND (Tools OR Libraries) AND (Fairness OR Responsible OR Privacy OR Auditing OR Transparency OR Security OR Explainability OR Interpretability). This adjustment was made to ensure the inclusion of more general tool-related terms alongside responsible AI terms, aiming to capture a broader range of open-source tools that may not be explicitly labeled under academic terms but are still relevant. Recognizing that Google search engine is designed to rank the most relevant and authoritative sources higher, we focused on extracting responsible ML libraries from the first 100 unique websites to capture a broader range of open-source tools. Similar approach has been used by previous studies to capture a wide spectrum of grey literature for their study \citep{washizaki2019studying, yasin2020using}. Following this, we compiled a list of open-source responsible ML libraries hosted on GitHub. From this Google search, we were able to identify a total of 35 responsible ML libraries. 
\item \textbf{Filtering responsible ML libraries}: The purpose of this step is to curate a refined list of active and relevant responsible ML libraries that are currently being developed and widely used within the community. After extracting the names of responsible ML libraries from different sources, we combined the lists (47) and removed duplicates (10), resulting in a final set of 37 libraries. Since we aimed to evaluate active and relevant ML libraries, we conducted an additional analysis collecting descriptive information available on each GitHub\footnote{\url{https://github.com/}} libraries repository page. Then, we assessed the number of stars, contributors, last commit, and programming language. Next, we removed libraries with less than 100 stars or a single contributor, as well as those whose last commit was dated before 2020 (regarding project activity history) to focus on libraries that demonstrate ongoing development and community engagement, as suggested by recent studies \citep{foalem2023studying,openja2022studying}.
Furthermore, we chose to focus exclusively on libraries developed for Python, given its prominence and widespread adoption in the machine learning field. This decision was informed by the language's robust ecosystem and the extensive use of Python among ML practitioners, which ensures that our findings are applicable to the majority of current ML projects \citep{hamidi2021towards}. Additionally, One author manually verified each library to ensure that it is an authentic responsible ML library, by checking the repository description and the official documentation. This process resulted in a list of 20 high-quality responsible ML libraries, described in \autoref{table:reponsiblelibrairies}.

\item \textbf{Collecting GitHub projects}: The purpose of this step is to compile a list of real-world projects that incorporate open-source responsible ML libraries. This aims to gather insights into the practical applications and adoption of these libraries across various domains.

For each previously selected library, we examined its GitHub repository, documentation, and associated source code to understand how the libraries might be used in practice. As a result, we gathered various configurations for using each library, such as \texttt{import captum}  and \texttt{from captum}, which we found for \texttt{captum}\footnote{\url{https://github.com/pytorch/captum}} on its GitHub page.
Using the import configurations as keywords, we conducted a code search through the GitHub search API to find repositories and corresponding source code using each responsible ML library. For instance, we searched for the keywords \texttt{import eli5} and \texttt{from eli5} to find repositories and source code using the \texttt{eli5}\footnote{\url{https://github.com/eli5-org/eli5}} responsible ML libraries. The full list of search queries is available in our replication package ~\citep{replication}.
We collected a list of 43,722 source code files that contained at least one import configuration of responsible ML libraries, along with the names of the corresponding repositories for all responsible libraries. 
Next, we eliminated duplicate repositories, resulting in 215 distinct repositories. To discard \emph{toy} projects from our sample, such as tutorial and student assignments repositories, we set a threshold of at least 50 commits and one star for each selected project. Additionally, one author manually reviewed all 85 projects to ensure their relevance and to exclude any remaining toy or low-quality projects.  After this thorough review process, all 85 projects were judged to be relevant and retained for our study \citep{replication}. Figure \ref{fig:overview} shows some descriptive metrics (number of stars, commits, contributors, and lines of code), collected for each project; providing an overview of our 85 selected projects.

\item \textbf{ML Dataset \citep{gonzalez2020state}}: The primary mean of this phase was twofold: (1) to gather a dataset of machine learning projects written in Python and relevant for our study, and (2) to analyze the issue discussions within these projects to identify discussions related to responsible AI principles. This targeted approach allowed us to focus our research on projects relevant to machine learning and likely to demonstrate engagement with responsible AI principles.
In this phase, we refined the publicly available dataset from \citep{gonzalez2020state} to select ML projects that aligned with our specific criteria: the projects must be developed in Python and must be ML-based applications. This filtering process resulted in a subset of 4,787 ML projects that met our requirements. From these projects, we extracted 44,877 issues directly from GitHub using the GitHub API. The purpose of analyzing a random sample of these issues was to gain insights into practitioners' discussions related to responsible AI principles.
By manually analyzing this sample, we aimed to identify whether responsible AI principles, such as fairness, privacy, transparency, and explainability, were explicitly or implicitly discussed. This step was critical in understanding how practitioners perceive and address responsible AI principles in their workflows, particularly within the natural context of issue discussions on GitHub.

\item \textbf{Survey Developers}: The purpose of this step was to directly engage with practitioners involved in machine learning development to assess their awareness, application, and perception of responsible AI principles. This direct feedback is crucial for validating our logging framework to build responsible AI applications. The survey was sent to contributors of the 85 analyzed ML projects, as these projects explicitly demonstrate the use of responsible AI principles in their implementation. In total, we gathered 2,735 email addresses from contributors.
To streamline communication, we developed a Python script to automate the process of sending emails to these practitioners. Our objective was to gather insights about practitioners' concerns regarding responsible AI principles and to validate our taxonomy. For this purpose, we designed a survey template comprising four sections: (1) Background Information: We collected data related to the respondents' roles and experience in ML/AI. (2) Awareness and Implementation of Responsible AI: In this section, we inquired whether practitioners were familiar with the concepts of responsible AI, if they currently implement these principles, and how logging practices could support these principles. (3) Responsible AI Principles: We asked which principles practitioners currently implement in their projects and how they determine the importance of specific responsible AI principles for their projects. The final section aimed to gather practitioners' concerns regarding our proposed taxonomy.
The survey was conducted from 18th February 2024 to 5th April 2024. We received 22 responses, resulting in a participation rate of 0.8\%, which is in line with
many software engineering surveys conducted outside specific companies \citep{uddin2019understanding, treude2015summarizing,nardone2023video}. The participants' backgrounds spanned both academia and industry, with 59.1\% being researchers, 27.3\% ML engineers, and 13.6\% data scientists. Regarding their experience in ML/AI, 31.8\% of the participants had more than 6 years of experience, 31.8\% had between 4 and 6 years, and 36.4\% had between 1 and 3 years. Notably, none had less than a year of experience. When queried about their familiarity with the concept of responsible AI principles, the responses were telling: 
18.2\% were not familiar with the concept at all, a sizable 45.5\% indicated they were somewhat familiar, 22.7\% felt they were familiar and engaged, and 13.6\% considered themselves very familiar. This spread of familiarity underscores the varying levels of awareness and engagement with responsible AI principles across the professional spectrum of our participants, highlighting both opportunities and challenges in enhancing the understanding and implementation of responsible AI practices within the field.
\end{enumerate} 
The subsequent data analysis will be outlined when discussing the results in Section \ref{sec:result}. We encourage replications of our study, and our replication package is available for the community \citep{replication}.
\begin{table*}[]
  \centering
  \caption{ \textbf{Summary of responsible ML libraries we investigated in ML-based applications. \\} \textit{ 
  Stars: Total number of stars. Contributors: Total number of contributors.  Size (KB):
Size of the tool. Releases: The total number of releases. Import configuration: import statement of the library.}}
  \begin{adjustbox}{width=\textwidth}
    \begin{tabular}{|c|l|r|r|r|r|l|}
    \toprule
    \multicolumn{1}{|l|}{\textbf{Responsible AI/ML principles}} & \textbf{Libraries} & \multicolumn{1}{l|}{\textbf{Stars}} & \multicolumn{1}{l|}{\textbf{Contributors}} & \multicolumn{1}{l|}{\textbf{Size}} & \multicolumn{1}{l|}{\textbf{Release}} & \textbf{Import configuration} \\
    \midrule
    \multirow{5}[10]{*}{Fairness} & facets & 7,115  & 31    & 23,279 & 4     & `from facets\_overview' \\
\cmidrule{2-7}          & FairML & 345   & 2     & 19,440 & 0     & `from fairml' \\
\cmidrule{2-7}          & Aequitas & 539   & 14    & 67,386 & 0     & from aequitas' \\
\cmidrule{2-7}          & AIF360 & 2,008  & 58    & 6,799  & 10    & `from aif360' \\
\cmidrule{2-7}          & fairlearn & 1547  & 75    & 77,953 & 12    & `from fairlearn' \\
    \midrule
    \multirow{3}[6]{*}{ Privacy} & tensorflow\_privacy & 1,754  & 50    & 3,065  & 14    & `from tensorflow\_privacy' \\
\cmidrule{2-7}          & opacus & 1,383  & 55    & 6,786  & 20    & `from opacus' \\
\cmidrule{2-7}          & crypten & 1,257  & 23    & 14,424 & 1     & `import crypten' \\
    \midrule
    \multirow{10}[20]{*}{ Transparency \& Explainability} & captum & 3,885  & 90    & 266,940 & 8     & `import captum', `from captum' \\
\cmidrule{2-7}          & yellowbrick & 3,978  & 112   & 80,675 & 24    & `from yellowbrick', `import yellowbrick' \\
\cmidrule{2-7}          & shapash & 2,112  & 23    & 60,832 & 23    & `from shapash'  \\
\cmidrule{2-7}          & shap  & 1,902  & 165   & 234,971 & 47    & `import shap' \\
\cmidrule{2-7}          & dowhy & 5,825  & 66    & 351,339 & 11    & `from dowhy', `import dowhy' \\
\cmidrule{2-7}          & eli5  & 2,616  & 15    & 37,429 & 0     & `import eli5' \\
\cmidrule{2-7}          & explain & 1,667  & 15    & 84,283 & 74    & `from explainerdashboard' \\
\cmidrule{2-7}          & dalex & 1,192  & 23    & 836,303 & 0     & `import dalex'  \\
\cmidrule{2-7}          & xai   & 897   & 3     & 18,684 & 1     & `import xai' \\
\cmidrule{2-7}          & aix360 & 1,319  & 23    & 367,188 & 2     & `from aix360' \\
    \midrule
    \multirow{2}[4]{*}{Safety \& Security} & syft  & 8,724  & 410   & 490,815 & 23    & `import syft' \\
\cmidrule{2-7}          & tf\_encrypted & 1,112  & 25    & 23,475 & 25    & `import tf\_encrypted' \\
    \bottomrule
    \end{tabular}%
    \end{adjustbox}
  \label{table:reponsiblelibrairies}%
\end{table*}%

\begin{figure}[htp!]
    \centering
    \includegraphics[width=0.6\textwidth]{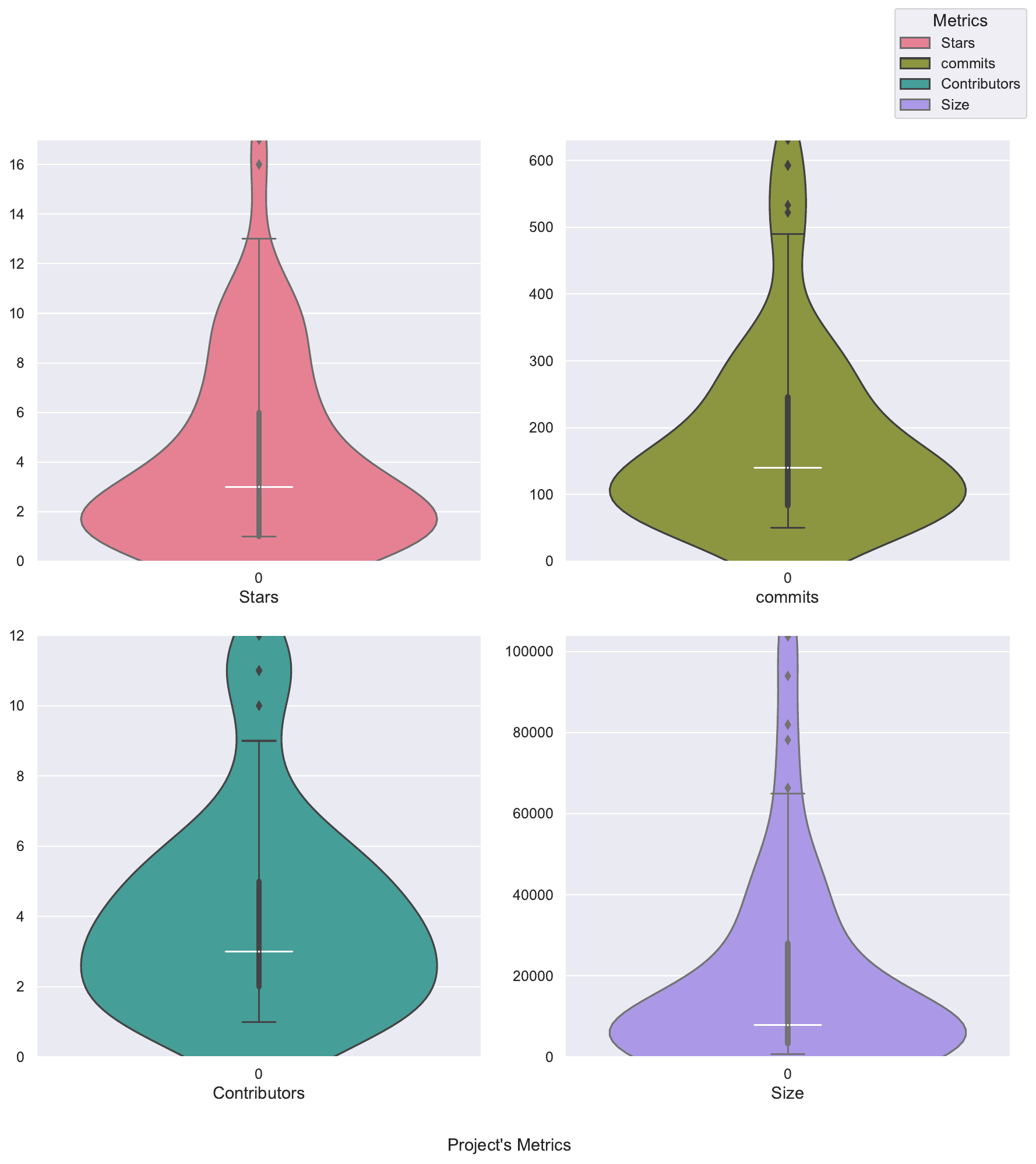}
    \caption{Overview of projects using responsible ML libraries selected in our study. \textit{ (5 percent of outliers are removed). The figure shows the number of contributors, commits, stars, and the project’s size (lines of code).}}
    \label{fig:overview}
\end{figure}
 
\section{Results}
\label{sec:result}
In this section, we present the results of our study. The raw results and the corresponding analysis are reported for each research question.

\subsection*{\textbf{RQ1: What responsible AI principles are considered by ML practitioners?}}
\begin{enumerate}[leftmargin=0.8 pt,align=left]
\item \textbf{Motivation:} 

Continuous auditing of ML applications requires a clear understanding of which responsible AI principles are actively considered by practitioners to ensure these systems adhere to ethical standards throughout their lifecycle. Responsible AI principles have been proposed as a framework for designing and deploying AI systems that are fair, transparent, and accountable \citep{de2021companies}. The identification of these principles is essential for developing effective auditing processes that can continuously evaluate compliance. By identifying the specific responsible AI principles that ML practitioners focus on, this research aims to highlight the areas where current measurement practices may be lacking and suggest ways to enhance the integration of these ethical considerations into ML development. This inquiry is pivotal for facilitating continuous and rigorous auditing mechanisms thereby advancing the creation of trustworthy and equitable AI systems.
\end{enumerate}
\begin{enumerate}[leftmargin=0.8 pt,align=left]
\item[2.]\textbf{Approach:}
\label{sec:subsectionapprocah}
To answer this research question, we adopted a mixed-methods approach that combines quantitative and qualitative analyses. We began by identifying libraries that support responsible AI principles following step in \autoref{sec:approach}, resulting in 20 libraries being selected (\autoref{table:reponsiblelibrairies}). We then conducted a detailed analysis of 85 ML projects on GitHub that utilize these libraries to provide a frequency analysis regarding their adoption. This was complemented by an analysis of issues collected from these ML projects and a survey of ML practitioners to complement our analysis. 

\noindent\textbf{Categorization of responsible AI libraries.} In our methodology to categorize the selected ML tools, we first analyzed each tool's official documentation, including the README file and repository description. This analysis was essential to understand the primary focus and application of each tool, as described by its developers. Based on this information, we created categories corresponding to responsible AI/ML principles. The categorization process was manual and involved two researchers examining the documentation to determine the most fitting responsible AI principle for each tool. This approach led to the development of principle categories derived directly from the tools' stated purposes and features. To ensure reliability in this classification, we measured inter-rater agreement, achieving a Cohen's kappa score of 0.71, indicating substantial agreement. In cases of disagreement, a third researcher was consulted, and discussions were held until a consensus was reached on the appropriate categorization. This careful and informed process ensured that each ML tool was classified into a responsible AI/ML principle category that accurately reflects its documented capabilities and intended use. 

\noindent\textbf{Extracting the use (function calls) of responsible AI libraries in GitHub ML projects.}
We then determined the most commonly implemented responsible AI principle among ML practitioners in our sample of selected projects. For that, we used an approach similar to \citet{majidi2022empirical}, where function calls of AutoML libraries were used to identify the usage of 
AutoML.
    We adopted the following steps for each project in our sample to extract and determine the number of function calls to quantitatively assess how extensively each responsible AI principle is being integrated into practical ML workflow:
    \begin{enumerate}[leftmargin=4 pt, align=left]
        \item[(i)] Extraction of all Python and Notebooks files.
        \item [(ii)] Conversion of all Python Notebook files to Python files using \texttt{nbformat} library \citep{nbformat2023}.
        \item [(iii)] Removal of all source code lines starting with specific expressions (\texttt{!, @, \%, \$, install, pip, coda}) in order to eliminate possible issues when parsing the file in the next steps. 
        \item [(iv)] Generation of AST for each file using the \texttt{AST} library \citep{PythonAST2023} from Python 3.
        \item [(v)] Extraction of all function calls based on the import configuration for each responsible AI library in the Python file, i.e., class, function, or simple statements related to responsible AI tools.         
     \end{enumerate}
\end{enumerate}

\noindent\textbf{Classification of function calls by the implemented responsible AI principles.}
After extracting 5,733 function calls, two authors manually classified each function call into one of the identified responsible AI principles. The classification process involved thoroughly reading and analyzing the official documentation related to the function calls of the respective libraries. This ensured that the classification aligned with the libraries' stated purposes regarding the function call. By referring to the official descriptions and/or use-case examples provided in the documentation, we ensured that each function call was appropriately associated with the responsible AI principle it was designed to address. This classification process achieved a kappa agreement of 0.81, indicating near-perfect agreement. Any conflicts in classification were resolved through discussions with a third researcher until a consensus was reached. We opted for manual classification over automation because libraries, such as Dalex, are not limited to a single responsible AI principle. For instance, while Dalex is primarily categorized under ``explainability", it also includes modules focused on fairness, as detailed in its documentation \footnote{\url{https://dalex.drwhy.ai/}}. This multifunctionality in libraries means that automated classification based solely on function calls might not accurately reflect all the responsible AI principles they support. We evaluate the number of function calls for each responsible AI principle to determine the frequency of their use in ML projects. Then, we grouped the results into four responsible AI principles based on the predominance of function calls related to import configuration and the stated purposes of each responsible AI library in their documentation. This grouping helped identify which principles are most actively implemented.

\noindent\textbf{Analysis of issues reports.} 
We conducted a random sampling of 382 reported issues from a pool of 44,877, achieving a 95\% confidence level with a 5\% confidence interval. Two researchers meticulously classified these 382 issues to determine whether they discuss responsible AI principles and, if they do, classified them into the respective 4 categories of responsible AI principles we identified earlier (Privacy, Transparency \& Explainability, Fairness, Security \& Safety).  The classification process involved carefully reading the issue titles and bodies to understand their context and content, and then assigning the identified responsible AI principles to the issues that matched one of these AI principle categories. This process attained a near-perfect kappa agreement of 0.82 across all categories. Any conflicts encountered during this process were thoroughly discussed in meetings until consensus was reached \citep{yamane1967statistics}. By analyzing these reported issues, we aimed to identify instances where responsible AI principles might have been a topic of discussion or concern among ML practitioners. This method of issue analysis was designed to systematically assess the prevalence and context of discussions on responsible AI within project issues. We hope that these findings will inform further research and encourage the adoption of best practices in the ML community.

\noindent\textbf{Survey practitioners:} We surveyed practitioners on the prevalent use of responsible AI principles in their projects to gather practitioners' insights and complement our analysis. We received 15 responses from practitioners divided between the 4 principles (\autoref{table:reponsiblelibrairies}). 

\begin{enumerate}[leftmargin=0.8 pt,align=left]
\item[3.] \textbf{Result:}
In this section, we present the findings of our investigation into the implementation of responsible AI principles by machine learning practitioners. Key findings include the identification of 20 libraries that support responsible AI principles and the prevalence of measurements of specific principles within ML projects, as indicated by the function calls in their code. Additionally, our analysis was enriched by examining a large subset of 44,877 reported issues related to ML projects and conducting practitioner surveys. The results collectively paint a comprehensive picture of what responsible AI principles are being integrated into real-world ML applications.
\item[3.1] \textit{Summary of Libraries Supporting Responsible AI Principles:} In our research, we identified and analyzed 20 libraries that support responsible AI/ML principles. These libraries, spanning the principles of Fairness, Privacy, Transparency \& Explainability, and Safety \& Security, represent a contribution to the field of AI ethics. a)~\textbf{Fairness:} The fairness category includes libraries such as Facets, FairML, Aequitas, AIF360, and Fairlearn. These tools, varying in popularity and contributor base, primarily focus on visualizing data for fairness analysis, auditing predictive models for bias, and providing metrics and algorithms to test and mitigate unfairness in ML models \citep{AIF3602023,facets2023,fairml2023,aequitas2023,fairlearn2023}. b) \textbf{Privacy:} In the realm of privacy, TensorFlow Privacy, Opacus, and CrypTen stand out. These libraries, backed by major tech entities like Google and Facebook, are instrumental in integrating differential privacy techniques and ensuring secure, private ML operations \citep{privacy2023,opacus2023,CrypTen2023}. c) \textbf{Transparency \& Explainability:} For transparency and explainability, libraries such as Captum, Yellowbrick, Shapash, SHAP, and DoWhy, along with others like ELI5, Explain, Dalex, XAI, and AIX360, offer a range of functionalities, including model interpretability, visualization for model diagnostics, making ML outputs understandable, and causal inference to comprehend model predictions \citep{captum2023,yellowbrick2023,shapash2023,shap2023,dowhy2023,DALEX2023,xai2023,AIX3602023,eli52023}. d) \textbf{Safety \& Security:} In safety and security, Syft and TF Encrypted are key players. Syft specializes in secure data sharing and federated learning, ensuring data privacy in distributed ML contexts. TF Encrypted extends TensorFlow's capabilities for secure computing, crucial for maintaining data confidentiality during ML processes. Both libraries underscore the importance of robust security measures in AI, vital for protecting data and model integrity. \citep{PySyft2023,encrypted2023}. Each library contributes to its respective principle, with varying degrees of community engagement and development activity, as indicated by their GitHub stars and the number of contributors. This diversity and breadth of tools highlight the growing focus and development in the field of responsible AI.
\item[3.2] \textit{Commonly Implemented Responsible AI Principles from 85 ML projects:} \autoref{table:reponsibleaiprinciplemetric} presents the total number of function calls for various responsible AI tools that ML practitioners currently consider when building ML-based applications; indicating priorities and focus areas of ML practitioners when integrating responsible AI principles into their ML-based applications. As seen in \autoref{table:reponsibleaiprinciplemetric}, ML practitioners frequently consider privacy 
when developing ML-based applications. This could reflect a growing awareness and concern for data protection and user privacy in the field of machine learning.
\begin{table}[htbp]
\centering
\caption{Overview of Metrics Across Responsible AI Principles}
\begin{adjustbox}{width=\textwidth,center}
\begin{tabular}{@{}lcccc@{}}
\toprule
\textbf{Metrics} & \textbf{Privacy} & \textbf{Transparency \& Explainability} & \textbf{Fairness} & \textbf{Security \& Safety} \\ \midrule
\# of function calls   & 5120 & 439  & 118  & 55   \\
\# of projects         & 26   & 47   & 14   & 3    \\
\# of issues           & 5    & 10   & 7    & 9    \\
\# of survey responses & 3    & 2    & 4    & 9    \\
\bottomrule
\end{tabular}
\end{adjustbox}
\label{table:reponsibleaiprinciplemetric}%
\end{table}

Regarding the number of projects implementing each principle, \autoref{table:reponsibleaiprinciplemetric} provides an insightful perspective on the distribution of responsible AI principles across the surveyed projects. We can observe that transparency \& explainability are implemented in over half of the projects, 52.22\% (47) projects, underscoring the importance of these principles in current ML practices, followed by privacy and fairness, suggesting a balance of concern for ethical considerations in ML development. However, the apparent lower emphasis on safety \& security raises questions about integration of responsible AI principles in ML projects.
   
Moreover, Figure \ref{fig:repos_by_number_of_principles} reveals that a minimal number of projects (only 5.9\%) incorporate more than one responsible AI principle. This indicates a potential gap in the comprehensive application of responsible AI principles in ML projects, suggesting that many projects may not be fully addressing the multifaceted ethical challenges in AI.

\begin{figure}[htp!]
    \centering
    \includegraphics[width=0.4\textwidth]{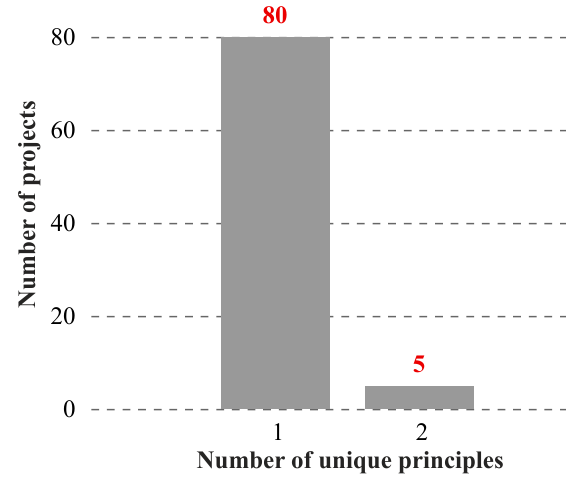}
    \caption{Number of responsible principles considered in ML project. \\ 
\textit{ \scriptsize {94.1\% of project audit only 1 aspect.}}}
    \label{fig:repos_by_number_of_principles}
\end{figure}
In summary, these results highlight a crucial area for improvement in the field of machine learning. While there is a clear acknowledgment of certain responsible AI principles like privacy and transparency \& explainability, the overall integration of these principles is not yet considered by ML practitioners. To build more responsible and trustworthy ML applications, practitioners should ensure that all four principles of responsible AI/ML are adequately addressed in their projects.
\item[3.3] \textit{Insights from Reported Issues and Practitioner Surveys:} 
We analyzed a random sample representing 0.85\% of the total 44,877 reported issues, which amounted to 382 issues. Out of these 382 issues, 31 issues—equivalent to 8.1\%—were found to include discussions on at least one responsible AI principles.
As presented in \autoref{table:reponsibleaiprinciplemetric}, Transparency \& Explainability emerged as the most frequently discussed principles among practitioners, followed by Security \& Safety, and then Fairness. Notably, Privacy was the least discussed principle.

From the survey of practitioners, we inquired about the responsible AI principles they most commonly implement in their ML projects. As presented in \autoref{table:reponsibleaiprinciplemetric}, Safety \& Security, and fairness emerged as the most considered principles. These were followed by Privacy, and then Transparency \& Explainability.
This trend indicates awareness and prioritization of Safety \& Security and Fairness among ML practitioners, suggesting that these aspects are increasingly viewed as critical components in the development of ethical AI systems. The relatively lower attention given to Transparency \& Explainability might reflect the nature of certain AI projects since not all applications require the same level of explanation or interpretability. For instance, ML projects involving less critical or sensitive decisions might not prioritize explainability, in comparison to 
more safety-critical domains such as healthcare or finance. 

    \begin{tcolorbox}[colback=black!4,colframe=black!50!white]

While privacy (89.3\%, 5,120 function calls) and transparency \& explainability (7.7\%, 439 function calls) dominate in implementation, security \& safety (1.0\%, 55 function calls) and fairness (2.1\%, 118 function calls) are significantly underrepresented despite being highlighted as crucial by 40.9\% of survey respondents. This reveals a gap between the principles practitioners value and those they implement, with no projects addressing all four responsible AI principles together.
    \end{tcolorbox}
\end{enumerate}
\subsection*{\textbf{RQ2: What runtime information of the ML applications is logged to support responsible AI auditing?}}
\begin{enumerate}[leftmargin=0.8 pt,align=left]
    \item \textbf{Motivation:} 
    Building on our findings from RQ1, which identified the key responsible AI principles that ML practitioners prioritize and the associated tools, RQ2 explores how these principles are actively considered and enforced through logging during runtime. The effective auditing of ML applications hinges on robust logging practices that capture critical runtime information aligned with responsible AI standards. This inquiry is crucial for ensuring that the operationalization of privacy, fairness, transparency, and other responsible AI principles is not only theoretical but also practical and traceable in real-world scenarios. By examining the specific data logged to support the use of responsible AI libraries, this research aims to shed light on the current state of logging practices and identify gaps where improvements are necessary. Enhancing these logging practices is fundamental for the rigorous auditing of ML applications; to verify their compliance with ethical standards and responsible AI guidelines. This deeper understanding will facilitate the continuous improvement of auditing mechanisms, making ML applications more trustworthy and ethical.
    
    \item \textbf{Approach:} 
    To answer this research question, we extract and analyze logging statements to assess their relevance to different responsible AI principles. Next, we survey developers to capture their perspectives on how their current logging practices support responsible AI.

    \noindent\textbf{Identification of logging libraries.} 
    To identify which information is collected via logging in ML projects, we extracted logging statements from our 85 study projects. There is currently limited related work discussing logging practices in ML projects from the practitioners' perspective. Therefore, we used a similar method proposed in the literature to identify ML logging libraries. Specifically, we conducted a Google search with the following query: (Machine Learning OR ML) AND (Logging) AND (Tools OR Libraries), examining the first 100 results pages, a similar approach and threshold have been used by \cite{foalem2024studying}. This allowed us to identify 20 relevant logging libraries. We then applied the steps outlined in the approach of our first research question (Subsection \ref{sec:subsectionapprocah}) for configuring the logging imports, which resulted in 7,558 logging statements (i.e., function calls to the logging libraries). To ensure accuracy in our data, we sanitized the logging statement dataset by removing all configurations such as \texttt{logging.getLogger}, \texttt{logging.disable}, \texttt{logging.config}, \texttt{wandb.init}, \texttt{logging.basicConfig}, etc., to eliminate false positives. This process left us with a total of 2,937 relevant logging statements.
    
    \noindent\textbf{Categorization of logging statements.}  The purpose of the logging categorization process was to identify and classify the logged metrics present in responsible ML projects, with a focus on metrics related to responsible AI principles or those that could be useful for auditing ML applications. This effort aimed to systematically uncover which metrics were being logged and evaluate their relevance to responsible AI principles, highlighting potential gaps and areas for improvement in logging practices.
    We performed stratified random sampling of 340 logging statements out of 2,937, ensuring proportional representation across the identified logging libraries. This included 250 occurrences from Python logging, 50 from TensorBoard, 22 from MLflow, 7 from dllogger, and 11 from Weights \& Biases. The stratified approach ensured that our taxonomy accounted for the diversity of logging practices across various libraries while maintaining a 95\% confidence level with a 5\% confidence interval \citep{yamane1967statistics}.
    We began with an initial pilot phase by randomly sampling 100 logging statements out of 340 selected logging statements. During this pilot phase, two authors independently reviewed each logging statement, carefully examining the log text, variables, and the surrounding code and comment to identify useful information describing logged metrics. Initial labels were created for each logging statement to describe its purpose. These labels were iteratively refined, grouped, and categorized to form a draft taxonomy of metric-related logging purposes. Throughout this process, the authors traveled back and forth between the logging statements and categories, updating and reorganizing the taxonomy to ensure consistency and coverage. A third researcher, with expertise in software engineering and machine learning logging practices, acted as an arbitrator to resolve ambiguities and disagreements, ensuring that the categorization process adhered to agreed-upon principles. The arbitrator also reviewed and validated the final taxonomy.
    After establishing the taxonomy in the pilot phase, the two authors independently analyzed the remaining 240 logging statements. To assess the reliability of their classifications, we calculated Cohen’s Kappa metric, which measures the level of agreement between the two authors. The Cohen’s Kappa for inter-rater agreement was 0.82, indicating near-perfect agreement. Statements that did not clearly fit into the existing taxonomy were marked and later discussed collaboratively to determine whether a new category should be created. Discrepancies in categorization were resolved by introducing the third researcher, who reviewed the contested cases and made final decisions. This iterative process allowed us to systematically refine and finalize the taxonomy, ensuring a comprehensive classification of metric-related logging purposes.
    The final taxonomy reflects the broad spectrum of logged metrics relevant to responsible AI auditing, providing insights into how ML practitioners currently log information and where improvements are needed to better support responsible AI practices.
    
    \noindent\textbf{Survey of developers.} To complement our technical analysis with real-world insights, we surveyed developers involved in ML projects.
    We asked,  \textit{``Do your current logging practices support responsible AI principles?"} to practitioners.
    This inquiry aimed to gather qualitative insights on how logging practices are perceived and implemented in the context of responsible AI. We received responses from 12 practitioners. By combining the analysis of logging statements with practitioner perspectives, we sought to gain a holistic understanding of the role of current logging practice in responsible ML applications.  
    
    \begin{figure}[htp!]
        \centering
        \includegraphics[width=0.6\textwidth]{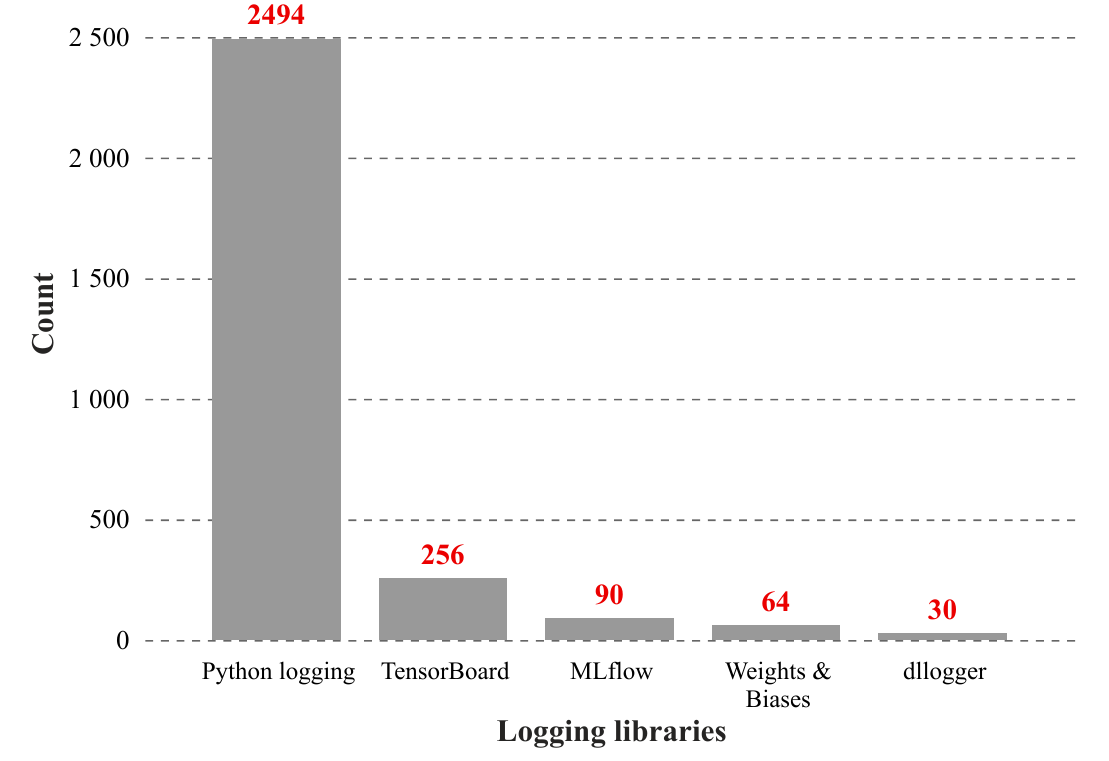}
        \caption{Logging libraries used in ML project.}
        \label{fig:logging_distribution}
    \end{figure}
   \item  \textbf{Results:} Figure \ref{fig:logging_distribution} illustrates the distribution of logging libraries across the machine learning projects we studied. As shown in the figure, ML practitioners predominantly use two types of logging libraries in ML projects: General Python libraries (such as logging) and ML-specific libraries (such as TensorBoard, MLflow, Weights \& Biases, and dllogger). The general Python libraries are the most commonly used logging libraries compared to the ML-specific libraries, which is consistent with the findings in \citet{foalem2024studying}. 

    \begin{figure}[htp!]
        \centering
        \includegraphics[width=0.6\textwidth]{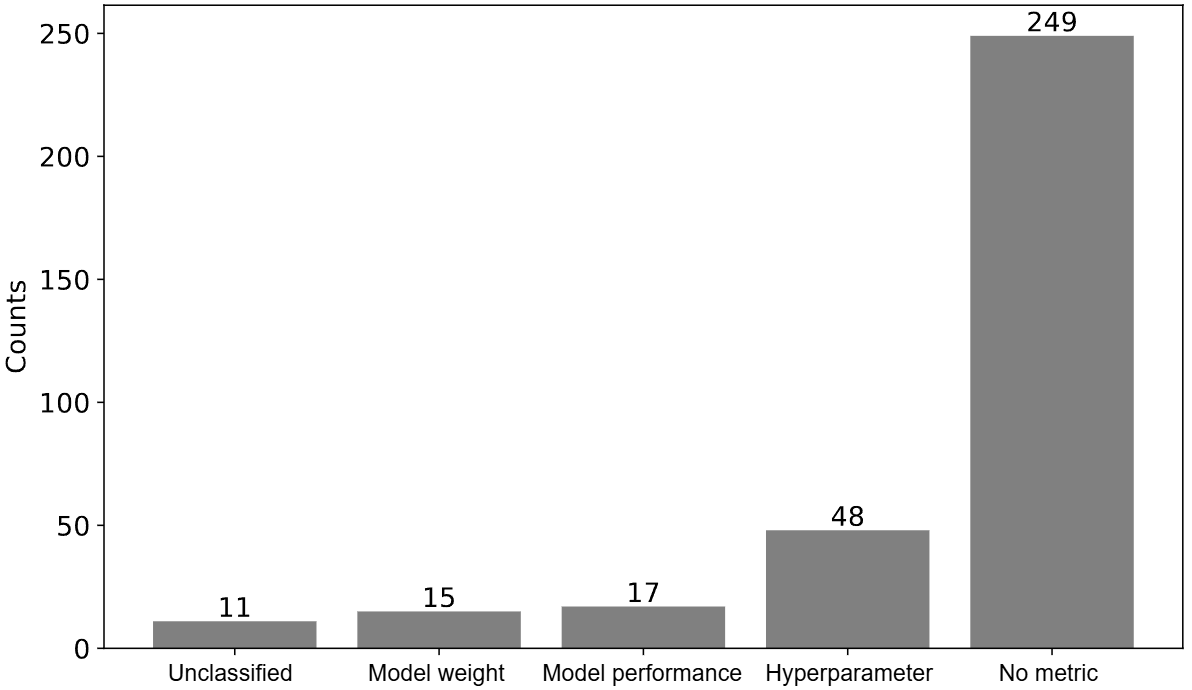}
        \caption{Logging statement classification.}
        \label{fig:logging_classification}
    \end{figure}
    
    Our categorization of logging statements revealed that out of the 340 sampled logging statements (\autoref{fig:logging_classification}), 14.1\% (48 statements) were used to log \textbf{hyperparameter} values in order  to track and reproduce the training process and its outcomes. These include settings such as learning rate, batch size, optimizer type, and regularization strength, which are essential for defining the behavior and performance of machine learning models. Similarly, 5.0\% (17 statements) were linked to \textbf{model performance} metrics, including accuracy, precision, recall, F1 score, and ROC-AUC score. These metrics are logged to assess the effectiveness and reliability of the model. Additionally, 4.4\% (15 statements) were used to log \textbf{model weights}, representing the learned parameters of the machine learning model that encapsulate the knowledge acquired during training and facilitate further analysis or retraining. A small portion, 3.2\% (11 statements), remained unclassified due to modifications made by contributors to the repository, making their purpose ambiguous.

    A significant proportion, 73.2\% (249 statements), were categorized as ``No metric." These statements primarily recorded general information, such as debugging messages or non-metric-specific details, rather than capturing specific metrics relevant to responsible AI principles. The taxonomy proposed by \cite{foalem2024studying} provides additional insights into the purpose of these ``No metric" statements, which often serve general debugging or operational monitoring needs in machine learning projects.
    The high percentage of ``No metric" logging statements highlights a critical gap in current practices, as these statements focus on general debugging purposes rather than logging data critical for auditing responsible AI principles.
    Practitioners provide a diverse range of insights regarding the question, \textit{Do your current logging practices support responsible AI principles?}
    Two researchers engaged in open coding of these responses to derive a clearer understanding of the underlying themes. The coding revealed six main categories reflecting the practitioners' views:
    
    \textbf{(1) Limited understanding and implementation}: Some practitioners showed a lack of understanding of how logging supports responsible AI principles, suggesting a need for better educational resources or guidelines.
    
    \textbf{(2) Awareness of principles but limited application}: While many practitioners are aware of responsible AI principles, monitoring these principles using logging remains minimal. For example, practitioner [P6] mentioned,  \includegraphics[width=2ex]{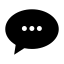} \textit{``Trying to work out if models I produce have any biases in them"}, highlighting awareness but not necessarily effective implementation.

    \textbf{(3) Intention to improve logging for responsible AI}: a respondent expressed a commitment to enhancing their logging practices in the future to better align with responsible AI standards. For instance: \includegraphics[width=2ex]{icons.png} \textit{"Not so much, unfortunately... But, we are also committed to best practices of audit, compliance, and responsible AI. So enhanced logging practices will definitely be something we look into and also implement in the near future."} -- [P18].

    \textbf{(4) Use of logging for specific responsible AI aspects}: A few practitioners explicitly mentioned using logging to support aspects like fairness and transparency or for debugging purposes related to AI ethics. For example, participants, [P6]and [16] said:\\ 
    \includegraphics[width=2ex]{icons.png} \textit{``Trying to work out if models I produce have any biases in them"} -- [P6]. \\
     \includegraphics[width=2ex]{icons.png} \textit{``It doesn't really. I log during training of course but not only to evaluate but also compare different models for class bias/imbalance."} -- [P16]. \\
    
    \textbf{(5) Disconnect between organizational commitment and practice}: a response indicated a gap between the organizational commitment to responsible AI and actual practice, with practitioner [P15] noting, \includegraphics[width=2ex]{icons.png} \textit{``In no way, because the top-down commitment is missing."}

    \textbf{(6) Practical use of logging unrelated to responsible AI}: Others noted using logging primarily for performance monitoring or debugging, not directly for supporting responsible AI principles, such as: \includegraphics[width=2ex]{icons.png} \textit{"My goal is to capture only system related events and data flow to fix any performance issues or check if there are potential data leaks."} -- [P22].\\
     \includegraphics[width=2ex]{icons.png} \textit{``I try to collect as much information as possible about the entire training process (model parameters, metrics, gradients, so on) in a parse-able format."} -- [P11] \\
    
    These results suggest that while there is an understanding of the importance of logging in ML projects, its application in the context of responsible AI principles is not yet fully realized. The varied responses indicate a potential gap in the integration of responsible AI considerations into current logging practices. This gap presents an opportunity for further development and refinement of logging strategies to more effectively support and enhance responsible AI principles in ML projects.
    \begin{tcolorbox}[colback=black!4,colframe=black!50!white]
    Overall, ML practitioners primarily log technical details like model performance, hyperparameters, and weights, with limited focus on metrics related to ethical considerations. This gap hinders the effective auditing of responsible AI tools. While practitioners recognize the importance of responsible AI principles, their current logging practices lack comprehensive coverage, highlighting an opportunity to improve logging for better ethical auditing of ML applications.
    \end{tcolorbox}  
\end{enumerate}

\subsection*{\textbf{RQ3: What specific logging requirements and information are essential for effective auditing of responsible AI principles?}}
\begin{enumerate}[leftmargin=0.8 pt,align=left]
    \item[1.] \textbf{Motivation:}
    Findings from RQ1 reveal that ML practitioners consider responsible AI principles such as fairness, privacy, and transparency in their applications. However, RQ2 highlights a critical gap: the information logged by these applications does not adequately reflect or support these principles. This discrepancy underscores the need for effective logging practices that bridge the gap between the responsible AI principles identified in RQ1 and the practical logging practices observed in RQ2. Possible reasons for this discrepancy could include limited awareness of the importance of comprehensive logging for auditing responsible ML applications, or a lack of clear guidelines on what specific information to log for responsible AI purposes.
    Effective improvement of current logging practices for auditing responsible machine learning applications requires identifying the relevant logging requirements or information that needs to be considered. This research question aims to identify key metrics or information that should be logged to provide valuable insights for auditing ML models. Without a clear understanding of the logging requirements or information, it may be challenging to identify potential issues in the data or the model, which can compromise the transparency, fairness, and privacy of ML applications. Therefore, this research question is essential to ensure that logging is integrated effectively into auditing practices for developing responsible and trustworthy ML applications.
    
    \item[2.] \textbf{Approach:}
    To address this research question, we followed a structured methodology combining insights from both RQ1 and RQ2 to develop a comprehensive logging framework for responsible AI.
    First, we began by analyzing function calls related to responsible AI principles extracted from our 85 study projects (RQ1). These function calls provided a foundation for identifying specific metrics and information essential for auditing responsible ML applications. Through manual analysis, we categorized the metrics into the four responsible AI principles identified in RQ1: fairness, privacy, transparency \& explainability, and reliability \& safety. Each metric was linked to a specific auditing purpose. For example, fairness metrics, such as group fairness and individual fairness scores, help evaluate whether models treat different demographic groups equitably, while privacy metrics, like epsilon and delta from differential privacy, assess compliance with data protection requirements.
    Using this categorization, we constructed the first half of our logging framework, which focuses on systematically capturing metrics aligned with responsible AI principles. This includes fairness scores (e.g., disparate impact ratio), privacy measures (e.g., epsilon and delta for differential privacy), and explainability indicators (e.g., SHAP values, feature importance). 
    Next, we incorporated findings from RQ2, where we added metrics currently logged by practitioners. These performance-related metrics, such as model accuracy, loss values, and hyperparameters, were added to the framework to complement the responsible AI metrics, ensuring that both responsible AI compliance and technical performance were adequately addressed.
    Together, these components formed a comprehensive logging framework designed to guide ML practitioners in systematically collecting and monitoring relevant information for responsible AI auditing. To validate the framework's applicability and relevance, we conducted a survey with ML practitioners, gathering insights on its practicality and alignment with real-world needs.
    
    \textbf{Identify relevant metrics.} We employed a stratified random sampling approach, preparing our dataset for qualitative analysis in step 7 of Figure \ref{fig:researchprocess}. Our sampling approach ensured a proportional representation of the four responsible principles identified in RQ1 in our sample. We randomly sampled 361 function calls, which provided a 95\% confidence level with a 5\% confidence interval \citep{yamane1967statistics}.
    To identify potential information or metrics that need to be logged for auditing responsible ML applications, we manually analyzed the official documentation for each function call in our sample. For example, by examining the functions  \textit{`aif360.sklearn.metrics.disparate\_impact\_ratio'} and \textit{`aif360.sklearn.metrics.
    consistency\_score'} of AIF360, we verified that they are used for evaluating fairness \citep{AIF360ConsistencyScore2023,AIF360DisparateImpactRatio2023}. This manual analysis involved two researchers achieving a kappa agreement of 0.71, i.e., substantial agreement, and any conflicts were resolved through discussions with a third researcher until reaching a consensus. This analysis helped us to identify any gaps in the relevant information or metrics that need to be considered and logged for improving current logging practices to build responsible and auditable ML applications.
    
    \textbf{Building a logging framework.}  We organized metrics identified previously into logging categories aligned with each responsible AI principle, such as fairness, transparency \& explainability, privacy, reliability \& safety. We also integrated performance-related logging information as outlined in RQ2, creating a comprehensive framework, since those performance-related metrics are currently logged by ML practitioners. 
    
    \textbf{Survey ML practitioners.} To further validate our proposed logging framework, we surveyed ML practitioners involved in real-world ML projects. We asked several questions to assess their agreement with the essential aspects of our framework, aiming to capture their insights on how well our proposed methods align with practical implementations. This survey helped confirm the relevance and applicability of our framework in supporting responsible AI principles through enhanced logging practices. 

    \item[3.] \textbf{Results:} Based on the detailed survey responses and the qualitative analysis of function calls related to responsible AI principles, we have developed a comprehensive logging framework as illustrated in Figure \ref{fig:Logging_diagram_cropped}. This framework is designed to enhance the auditability of ML applications by ensuring they adhere to responsible AI principles. It specifies metrics and information that should be continuously monitored and logged to support auditing processes.\\
    The logging framework integrates specific metrics identified during the analysis, related to the 4 principles we previously identified, which are essential for upholding responsible AI principles. These metrics are categorized into distinct logging categories corresponding to each responsible AI principle, Fairness, Transparency \& Explainability, Privacy, and Reliability \& Safety. Each category addresses predefined audit questions, guiding ML practitioners on what information should be logged to ensure their applications are both responsible and auditable.\\
    This structured approach helps to clarify how each aspect of the model's operation contributes to overall accountability and ethical compliance. By aligning logging practices with these detailed categories, the framework aims to facilitate a more thorough and efficient audit process, enabling practitioners to detect and address issues proactively. The framework encourages continuous monitoring and updating of logging practices to address emerging privacy, security, and ethical concerns effectively, thereby enhancing the trustworthiness and reliability of ML applications.
    This integrated approach not only aligns with the insights from our qualitative analysis but also reflects the broad consensus among practitioners about the importance of structured and comprehensive logging practices to support responsible AI auditing. The framework, therefore, serves as a crucial tool for practitioners, helping them to implement logging practices that are robust, systematic, and aligned with the core principles of responsible AI.\\
    \begin{figure}
        \centering
        \captionsetup{justification=centering,margin=2cm}
        \includegraphics[width=0.9\textwidth]{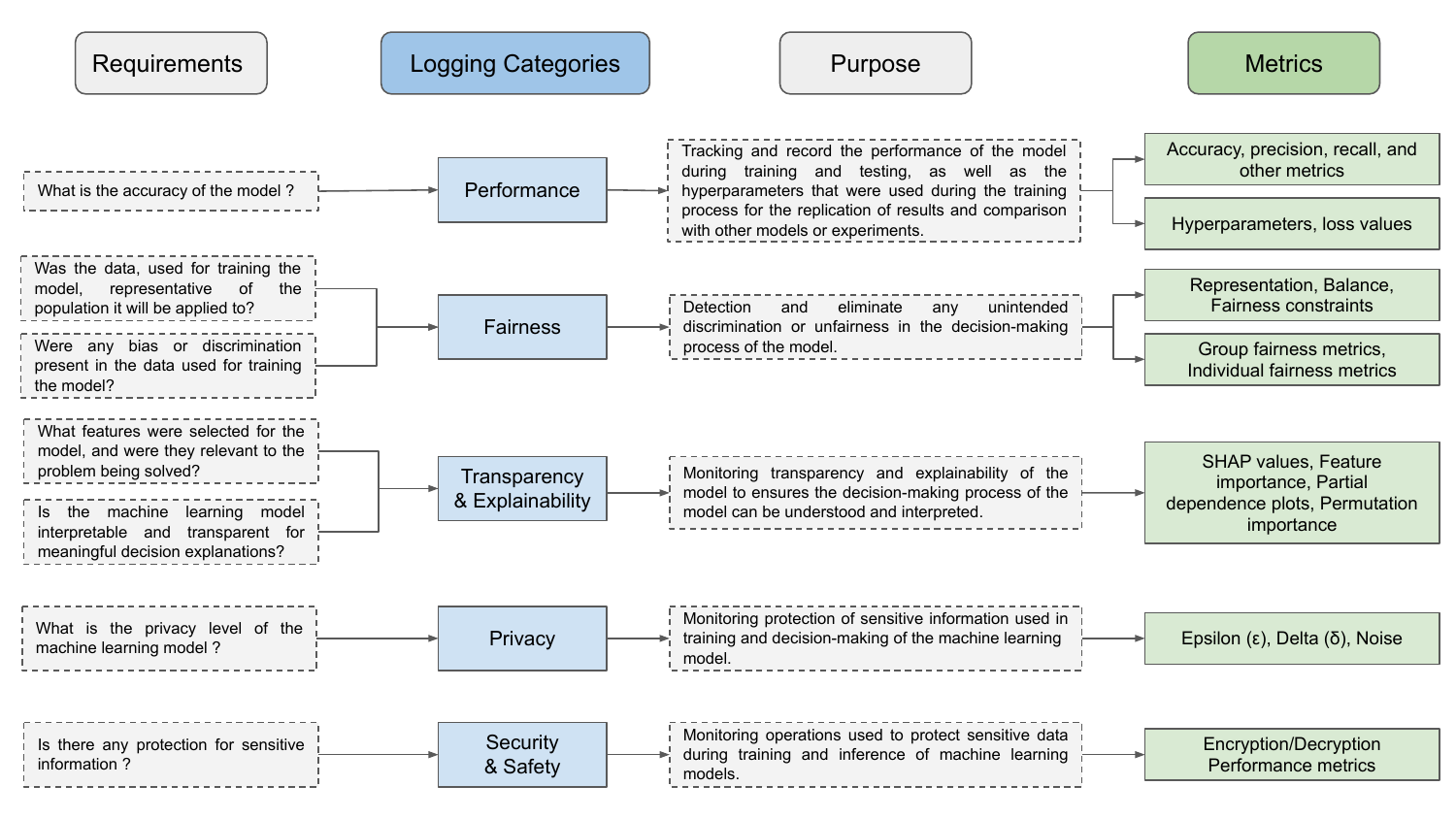}
        \caption{Logging Framework for responsible ML applications. \\ 
    \textit{ \scriptsize {For each requirement (left side), we associate a logging category, explaining its purpose and metrics that can be adopted to evaluate compliance with each principle (right side).}}}
        \label{fig:Logging_diagram_cropped}
    \end{figure}

    \item[3.1]  \textit{Logging categories and associated metrics:} we discuss in this section the purpose and metrics of each logging category present in our logging framework.\\ 
    
\textbf{$\bullet$ Performance:} In this category, ML practitioners usually log two types of information: (i) model training and validation, and (ii) business goal. Next, we discuss how each type of information might be obtained by a set of related metrics. For the first category, the required information is critical in tracking the progress of the model during the training phase. The information that should be logged includes \textbf{hyperparameters} used in training, such as learning rate, optimizer, batch size, and number of epochs. These parameters are essential for reproducing the training process and help in tuning the model for better performance. The second category refers to the key performance indicators that are used to measure the success or effectiveness of a machine learning model in achieving business goals. These metrics can vary depending on the specific business problem and industry, but some common examples include; accuracy, precision, recall, and loss value. Other metrics, such as F1 score, area under the curve (AUC), and confusion matrix, can also be logged to provide a comprehensive overview of the model's performance. By logging these metrics, practitioners can monitor the progress of the training process and make necessary adjustments to the model architecture or training data.\\

\textbf{$\bullet$ Fairness:} In this group, we observed that ML practitioners evaluated various fairness metrics but did not log them. For example, in project credit-risk\footnote{\url{https://shorturl.at/tANVX}}, we observed that practitioners saved fairness metrics in a CSV file. However, we suggest that these fairness metrics should be logged in (i) data and (ii) model training perspective. Next, we discuss how each perspective might be addressed by a set of related metrics. 
For the data processing perspective, we observe the use of the following metrics: a) \textbf{Representation}: An essential aspect of fairness is ensuring that the training data is representative of the population being modeled. Here, practitioners compute the proportion of the dataset that belongs to different demographic groups (e.g., race, gender, age) \citep{pessach2022review}. b) \textbf{Balance}: In addition to representation, it is also important to ensure that the training data distribution is balanced across demographic groups. Here, practitioners compute the difference between the proportion of positive examples in the dataset for each demographic group \citep{pessach2022review}. c) \textbf{Fairness constraints}: In some cases, practitioners enforce fairness constraints on the training data during the preprocessing step and try to mitigate bias in the processing step before the model training. For example, one may want to ensure that the difference in acceptance rates between different demographic groups is below a certain threshold \citep{pessach2022review}.  

For the model training perspective, we observe the use of the following metrics: a) \textbf{Group fairness}: the set of metrics used to assess whether a machine learning model is fair with respect to different demographic groups, such as race or gender. These metrics compare the performance of the model across different groups and identify disparities in the model's predictions. Some common group fairness metrics include equalized odds, equal opportunity, and demographic parity \citep{wan2021modeling}. b) \textbf{Individual fairness}: the set of metrics that aim to ensure that similar individuals are treated similarly by a machine learning model \citep{wan2021modeling}. These metrics focus on measuring the similarity between individuals based on their features and how the model treats them. Some common individual fairness metrics used by practitioners include Disparate Impact in Similarity, Generalized Entropy Index.     
Overall, logging fairness metrics at both stages of the ML pipeline can help in identifying and mitigating potential biases and ensuring that the model's decisions are fair and unbiased.\\

\textbf{$\bullet$ Transparency \& Explainability:} in this category, similar to fairness, we observed that ML practitioners do not log transparency and explainability metrics. These metrics include: a) \textbf{SHAP values}: These are a measure of feature importance that shows how much each feature contributes to the model's prediction for a given data point. They can be logged for each prediction and used to identify which features are most influential in the model's decision-making process \citep{parr2024nonparametric,yang2022interpretability}.
b) \textbf{Feature importance}: This is another measure of feature importance that can be used to identify which features have the greatest impact on the model's predictions. It can be logged for each feature and used to prioritize feature engineering efforts or to identify potential sources of bias or error \citep{parr2024nonparametric,yang2022interpretability}. c) \textbf{Partial dependence plots}: These are visualizations that show how the model's predictions change as a function of one or more features while holding all other features constant. Information related to this visual can be logged for each feature and used to gain insights into the relationship between the features and the model's predictions \citep{greenwell2020variable,yang2022interpretability}. d) \textbf{Permutation importance}: This is a method for measuring feature importance by randomly permuting the values of each feature and measuring the effect on the model's performance. It can be logged for each feature and used to validate the model's feature selection or to identify potential sources of overfitting \citep{greenwell2020variable}.      
By logging these metrics, practitioners can ensure that their model is transparent and explainable, allowing for better interpretability and understanding of the model's behavior.\\

\textbf{$\bullet$ Privacy:} In this group, we also observe that practitioners do not log metrics regarding privacy. Next, we present metrics that cover this current responsible AI principle: a) \textbf{Epsilon ($\epsilon$)}: is a measure of the privacy loss associated with differential privacy. It represents the maximum amount that an individual's privacy can be compromised by the use of the algorithm. Lower values of $\epsilon$ correspond to higher levels of privacy \citep{dwork2019differential}. b) \textbf{Delta ($\delta$)} is another measure of the privacy loss associated with differential privacy. It represents the probability that an individual's privacy will be compromised by the use of the algorithm. Lower values of $\delta$ correspond to higher levels of privacy \citep{arachchige2019local}. c) \textbf{Noise}: Many privacy-preserving techniques, such as differential privacy, add noise to the data or model to protect privacy. The amount and type of noise being added may be logged for analysis and debugging purposes. Logging privacy metrics is crucial for ML practitioners to ensure that they are complying with ethical and legal standards related to privacy. Additionally, logging privacy metrics can help to build trust with stakeholders, as it demonstrates a commitment to protecting sensitive information \citep{geng2015optimal}.\\

\textbf{$\bullet$ Security \& Safety:} Our observation shows that ML practitioners use encryption/decryption techniques such as Secure Multi-Party Computation (MPC), Homomorphic Encryption (HE), and Secret Sharing to protect sensitive data, but they do not log any information related to it \citep{yang2019comprehensive, xu2022privacy}. Therefore, we recommend that ML practitioners log \textbf{encryption/decryption performance efficiency} to track the time it takes for encryption/decryption operations to be completed on a particular dataset. By recording this information, they can calculate the average or median time for encryption/decryption and compare it against a predefined threshold to determine if the performance meets the required level. ML practitioners should monitor encryption/decryption performance efficiency to evaluate the efficiency and effectiveness of encryption/decryption operations in encrypted machine learning systems. Additionally, they can log information such as the \textbf{encryption/decryption technique} used to provide insights into the security and safety of the encryption/decryption process. By doing so, they can ensure the security and safety of the system, and detect any potential issues or inefficiencies in the encryption/decryption process. \\


\item[3.2]  \textit{Framework validation with ML practitioners:} Following our detailed proposition of a logging diagram to support the auditing process of ML applications in alignment with responsible AI principles, we sought to validate this framework through a practitioner survey. The survey aimed to gauge practitioners' agreement with the essential aspects of logging identified in our diagram, and to collect their input on any additional information that could enhance the auditing process. 
To explicitly measure the practitioners' agreement on both the logging principles and the specific metrics suggested in our framework, we presented the framework (\autoref{fig:Logging_diagram_cropped}) to them and posed several targeted questions. Each question was designed to assess their agreement on critical aspects of logging for auditing purposes. These included:
\begin{itemize}
    \item[$\bullet$] \textit{Do you agree that logging performance metrics, including model training and testing outcomes and hyperparameters, is essential for auditing the replicability and comparability of AI models?}
    \item[$\bullet$] \textit{Do you agree that logging for fairness, specifically detecting and mitigating unintended discrimination in model decision-making, is vital for auditing responsible AI applications?}
    \item[$\bullet$] \textit{Do you agree that ensuring the transparency and explainability of AI models through logging is crucial for the auditing process to validate the model’s decision-making?}
    \item[$\bullet$] \textit{Do you agree that the auditing of responsible AI applications should include logging practices that monitor and protect sensitive information during training and decision-making processes?}
    \item[$\bullet$] \textit{Do you agree that for auditing reliability and safety in AI models, logging the methods used to protect sensitive data during training and inference is necessary?}
\end{itemize}
Additionally, we asked practitioners to suggest any other metrics or information they felt should be included in the logging framework to further enhance its effectiveness and relevance. \\

The responses to these questions provided us with direct feedback on both the conceptual agreement with the logging principles and the practitioners' acceptance of the specific metrics we proposed for each category within our logging framework.

\begin{table}[htbp]
\centering
\caption{Overview of Practitioner Survey Results on Validating Framework}
\label{tab:survey_validation}
\begin{adjustbox}{width=\textwidth, center}
\small 
\begin{tabular}{@{}>{\raggedright\arraybackslash}p{3.5cm}cccccc@{}}
\toprule
\textbf{Responsible AI Principle} & \textbf{Strongly Agree} & \textbf{Agree} & \textbf{Disagree} & \textbf{Strongly Disagree} & \textbf{Total Responses} & \textbf{Agreement Percentage} \\ \midrule
Performance Metrics & 14 & 8 & 0 & 0 & 22 & 100\% \\
Fairness Metrics & 7 & 14 & 1 & 0 & 22 & 95.45\% \\
Transparency \&\newline Explainability & 5 & 14 & 3 & 0 & 22 & 86.36\% \\
Privacy Metrics & 7 & 12 & 3 & 0 & 22 & 86.36\% \\
Reliability and\newline Safety Metrics & 5 & 13 & 3 & 1 & 22 & 81.82\% \\
\bottomrule
\end{tabular}
\end{adjustbox}
\end{table}


As shown in the \autoref{tab:survey_validation}, practitioners universally acknowledged the importance of logging performance metrics, with 100\% agreeing that these metrics are critical for the replicability and comparability of AI models. This strong consensus validates the performance category of our proposed logging framework, emphasizing its essential role in responsible AI auditing.\\

The survey results also highlight the critical need for logging practices focused on fairness, transparency and explainability, and privacy. Notably, the feedback on fairness metrics was particularly strong, with nearly 95.45\% of practitioners endorsing the integration of these metrics into the auditing process. This reflects a strong consensus on the significance of fairness as an indispensable pillar of ethical AI development.\\

Moreover, a significant majority of practitioners agreed (86.36\%) on the importance of logging for transparency and explainability, as well as privacy metrics, which are crucial for maintaining the integrity and trustworthiness of ML applications. These responses affirm the necessity of including such metrics in responsible AI auditing frameworks.\\ 

Finally, 81.82\% of practitioners agreed on the importance of documenting the measures taken to secure sensitive data throughout the AI model lifecycle.\\

We further asked practitioners, ``Which of the metrics presented in our logging framework do you consider, document, or log when implementing one of the responsible AI principles?'' Figure \ref{fig:Metric_from_survey} shows the practitioner responses, highlighting a significant focus on conventional performance metrics such as accuracy, precision, recall, and similar metrics, which received the highest number of responses (20). This is followed closely by hyperparameters and loss values with 19 responses, indicating a strong emphasis on model performance and optimization aspects in logging practices.

However, other responsible AI metrics such as fairness constraints, group fairness metrics, and especially individual fairness metrics, which received no responses, show much lower consideration. This discrepancy suggests that while ML practitioners are keen on logging standard performance data, there remains a gap in the logging of metrics crucial for ensuring fairness and other ethical considerations.

Metrics related to privacy and security, like encryption/decryption performance metrics, are also among the least considered, with only one response; highlighting potential areas for improvement in the auditing and monitoring of ML applications for compliance with responsible AI principles.

Overall, the survey validated our proposed logging framework and indicated a strong alignment with the participants' views on responsible logging practices. The feedback not only validates the selected metrics within each logging category but also suggests areas for expansion, such as including metrics for environmental impact, energy consumption, and ethical considerations of AI applications to build responsible and trustworthy ML applications, as suggested by some ML practitioners. While practitioners expressed positivity about the proposed framework (Table \ref{tab:survey_validation}), the lack of consideration or recording of metrics (Figure \ref{fig:Metric_from_survey}) stems from several challenges highlighted during the survey (RQ2). These challenges include limited awareness of the importance of logging specific responsible AI metrics, technical barriers in implementing comprehensive logging practices, and competing priorities that may deprioritize logging in favour of other methods, such as testing or external audits. Addressing these challenges will be crucial to bridging the gap between responsible AI principles and their practical implementation in logging practices.

\begin{figure*}
    \centering
    \includegraphics[width=0.99\textwidth]{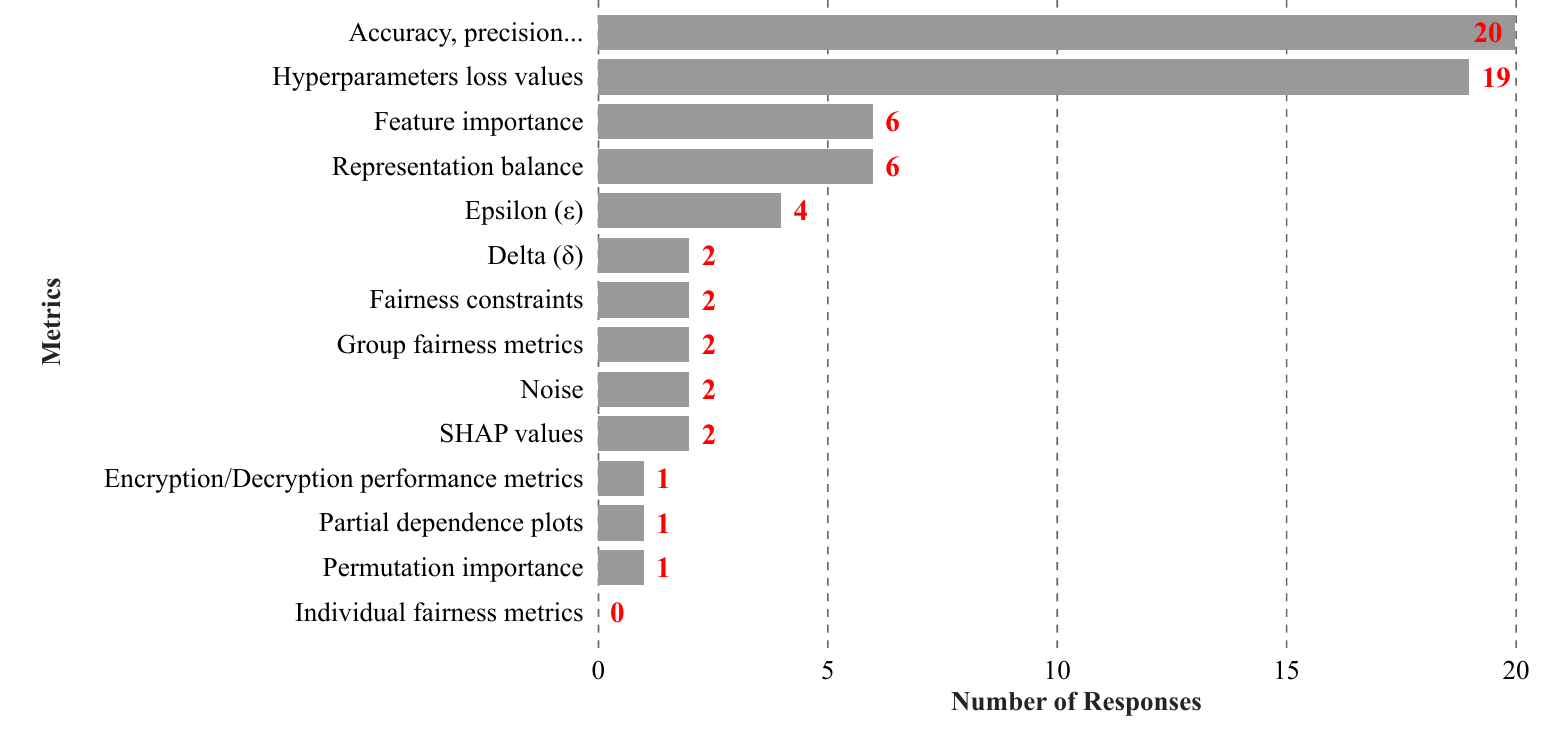}
    \caption{Practitioners overview on different logging metrics}
    \label{fig:Metric_from_survey}
\end{figure*}

\begin{tcolorbox}[colback=black!4,colframe=black!50!white]
To develop responsible and auditable ML applications, practitioners are suggested to log information aligned with responsible AI principles. Our study led to the creation of a logging framework designed to address gaps by incorporating a broader range of responsible AI metrics. Validated by practitioners, this framework enhances auditing and accountability, ensuring ML applications adhere to ethical and legal standards while promoting trust and reliability.
\end{tcolorbox}
\end{enumerate}


\section{Discussion and implication}
\label{sec: discussion_implication}
In this section we discuss the implications of our findings for researchers, ML practitioners, and tool developers.
\begin{itemize}[leftmargin=0.8 pt,align=left]
    \item \textbf{Researchers}: Our study provides a snapshot of current practices among ML developers with respect to logging requirements for auditing responsible ML applications. Our findings reveal that out of the four responsible principles, ML practitioners mainly focus on two aspects: privacy and transparency \& explainability. Moreover, none of the projects we examined integrated all four responsible principles (Finding from RQ1). To address this gap, the research community can play a role in developing comprehensive and robust methods for auditing ML applications in order to improve their trustworthiness. This can involve investigating and proposing ways to incorporate all four responsible ML principles associated with no ML component properties  in the audit process. Developing guidelines, standards, and best practices for auditing machine learning applications that account for all four principles can also be a focus of the research community. 
    
    Regarding the integration of logging in auditing practices for ML-based applications, our investigation highlights the fact that ML practitioners tend to focus only on monitoring performance through logging rather than considering metrics and information related to responsible AI principles (Finding from RQ2). To address this issue, researchers can develop and propose standardized logging formats and guidelines for different categories of metrics, including performance and responsible AI principles. For instance, with respect to the fairness principle, researchers could propose logging guidelines that require capturing and recording group fairness metrics and individual fairness metrics.
    
    \item \textbf{ML practitioners}: Our study provides actionable insights into the current logging practices and highlights areas for improvement in auditing responsible ML applications. Our findings indicate that ML practitioners predominantly log only business metrics and ignore responsible AI metrics (Finding from RQ2). Practitioners should develop mechanisms to log responsible AI metrics such as fairness metrics (e.g., group and individual fairness scores) and privacy indicators (e.g., differential privacy parameters like epsilon and delta) at different stages of model training and deployment. By tracking how these metrics evolve over time, practitioners can identify trends that may signal potential issues, such as increasing bias in model outputs or weakening privacy guarantees. Based on our proposed framework in RQ3, practitioners should consider leveraging pre-built logging templates designed specifically to develop trustworthy and responsible ML-based systems.  They should also consider logging as a collaborative practice involving diverse stakeholders such as ethicists, legal experts, and domain specialists. Such collaboration can ensure that logging strategies align with organizational goals, regulatory requirements, and societal expectations. For example, input from legal teams can inform logging strategies that better document compliance with regulations like GDPR, while ethicists can guide the selection of critical fairness and transparency metrics to log based on societal impact and ethical considerations.

    
    \item \textbf{Logging Tool developers}:  Our findings present an opportunity for logging tool developers to innovate by addressing the specific needs of responsible AI auditing. Beyond automating the capture of existing responsible AI metrics, tool developers should focus on creating features that actively guide ML practitioners in identifying and logging relevant metrics for fairness, privacy, transparency, and security. For example, logging tools could integrate intelligent suggestions that recommend appropriate metrics to log based on the type of ML application, data characteristics, or domain-specific requirements. Such context-aware logging capabilities would provide practitioners with tailored guidance, making it easier to adopt responsible AI practices.

\end{itemize}

\section{Related works}
\label{sec:related_work}
In this section, we introduce and discuss related work to our study. Initially, we present studies regarding auditing in ML systems. Next, we discuss the findings of previous studies regarding responsible AI principles. 

\subsection{Audit in Machine Learning}
The audit of ML is an emerging field, with recent works introducing diverse methodologies and perspectives to tackle the complex challenges associated with auditing AI and ML applications. The Office of the Auditor General of Norway's initiative \citep{algorithmswhite} serves as a foundational guide for public auditors examining AI and ML technologies within public sectors. This seminal work highlights the importance of data quality, project management, and governance in AI systems, promoting transparency and accountability to mitigate risks such as automated discrimination and data security breaches. Moreover, in the healthcare domain, \citet{oala2021machine} stress the importance of rigorous auditing and quality control for Machine Learning for Health (ML4H) tools. They advocate for a comprehensive approach to algorithm auditing to improve the reliability and effectiveness of ML systems in healthcare settings.

\cite{mlinauding} further examines the transformative impact of ML on the auditing profession. By enhancing audit speed and quality through in-depth data analysis, this contribution underscores the shifting role of auditors in a data-centric audit environment. Fairness in ML auditing also garners attention \citep{saleiro2018aequitas, kearns2018preventing, yan2022active}, with \citet{park2022fairness} proposing a framework that addresses data privacy, model secrecy, and trustworthiness to ensure algorithm fairness. This framework employs confidential computing and a trust chain established through enclave attestation primitives, enabling the secure certification of fair ML models.

In the emerging field of foundation models, \citet{mokander2023auditing} highlights the challenges of auditing large language models (LLMs), proposing a structured approach that includes governance, model, and application audits, inspired by IT governance and systems engineering best practices. Despite their innovative approach, they acknowledge the difficulty in operationalizing concepts like robustness and truthfulness, which are crucial for the auditing process.

Our research distinctly targets the logging requirements essential for continuously auditing responsible ML-based applications. By focusing on this area, we aim to facilitate effective and efficient audits that ensure these applications meet ethical and legal standards, thus addressing a critical gap identified in the existing literature.

\subsection{Responsible AI principles}
The discussion of Responsible AI principles Fairness, Accountability, Transparency \& Explainability, Privacy, and Security \& Safety has gained significant attention in both research and industry. In this section, we provide a structured overview of related work, organized by each principle.

\textbf{Fairness} focuses on addressing biases and ensuring equitable outcomes in AI systems. Mehrabi et al. \citep{pessach2022review} conducted a comprehensive survey on bias and fairness in machine learning systems. They categorized sources of bias, real-world applications affected by bias, and methods to mitigate it. The authors provided a taxonomy of fairness definitions and emphasized the importance of tackling algorithmic bias to prevent discriminatory outcomes. Similarly, Holstein et al. \citep{mehrabi2021survey} conducted semi-structured interviews and surveys with 267 ML practitioners, identifying challenges faced in building fair ML systems. They highlighted gaps between research solutions and the needs of industry practitioners, such as the difficulty of integrating fairness tools into workflows. While these studies address fairness challenges, they primarily focus on bias mitigation tools and fail to discuss logging requirements for auditing fairness in ML projects.

\textbf{Accountability} in machine learning aims to assign responsibility for system decisions and behavior. Cooper et al. \citep{cooper2022accountability} revisited the concept of accountability in an algorithmic society. They examined barriers that machine learning systems pose for accountability and proposed relational frameworks to address these challenges. Their analysis emphasized the need for robust systems that clearly define roles and responsibilities among stakeholders. Similarly, Rakova et al. \citep{rakova2021responsible} explored practical challenges organizations face in implementing responsible AI, highlighting enablers such as leadership support and ethical frameworks. While these studies underscore the theoretical importance of accountability, they do not provide concrete technical solutions, such as how accountability metrics can be logged or measured during the system’s lifecycle.

\textbf{Transparency and explainability} focus on making AI systems interpretable and understandable for stakeholders. Arrieta et al. \citep{arrieta2020explainable} presented a survey of Explainable AI (XAI), highlighting methods for achieving transparency and the challenges associated with implementing interpretable models. Their work categorized techniques like SHAP and LIME as critical tools for understanding model behavior. Peters et al. \citep{peters2020responsible} emphasized embedding explainability frameworks into AI engineering practices, demonstrating their use in digital mental health applications to promote fairness and transparency. However, these studies primarily focus on explainability tools and techniques without considering their integration into logging mechanisms for continuous auditing.

\textbf{Privacy} addresses concerns related to safeguarding user data during ML operations. Liu et al. \citep{liu2021machine} surveyed privacy-preserving techniques in machine learning, such as differential privacy and homomorphic encryption. They categorized privacy solutions into three areas: private ML training, privacy-aided ML protection, and ML-based privacy attacks. abbas et al. \citep{abbas2022safety} provided a comprehensive review of privacy threats and solutions in ML, identifying challenges like balancing privacy and model utility. 

\textbf{Security and safety} focus on ensuring ML systems are robust against attacks and function reliably in real-world settings. Abbas et al. \citep{abbas2022safety} reviewed machine learning solutions for addressing cyber-physical security and privacy challenges in IoT systems. They highlighted threats like data breaches, spoofing, and denial-of-service attacks, emphasizing the need for dynamic and adaptive security measures.

Collectively, these studies show the multifaceted nature of responsible AI principles, encompassing technical, organizational, and ethical dimensions. They offer a roadmap for transitioning from high-level ethical principles to concrete practices, ensuring AI technologies are developed and deployed in a manner that respects and enhances human values. However, they do not specifically focus on the logging requirements for continuous auditing machine learning-based applications for responsible AI principles from a practical perspective. Our study brings a novel perspective to the discussion of responsible AI by examining the role of logging in auditing ML applications for trustworthiness.  

\subsection{Logging in Machine Learning}
The role of logging in machine learning (ML) applications has gained growing attention in recent years, particularly in the context of production ML pipelines. \cite{foalem2024studying} highlighted the distinction between general-purpose logging libraries, such as Python's logging module, and ML-specific logging tools like MLflow, TensorBoard, Neptune, and Weights \& Biases (W\&B) (mlflow, 2024; neptune, 2024; tensorboard, 2024). These tools primarily facilitate debugging, monitoring performance, and maintaining reproducibility in ML workflows but often fall short in addressing logging needs directly aligned with responsible AI principles, such as fairness, explainability, and privacy.

\cite{shankar2021towards} expanded on the critical role of observability in ML pipelines by proposing a "bolt-on" observability framework to support end-to-end monitoring, diagnosis, and response to issues in deployed ML systems. Their work emphasizes the importance of logging fine-grained data at the component level to diagnose pipeline errors, track data quality constraints, and detect data distribution shifts. These approaches align with the need for a comprehensive logging strategy to address silent failures and support responsible AI initiatives effectively.

Despite advancements in logging frameworks, current practices predominantly focus on capturing metrics for performance evaluation, such as accuracy or precision, while neglecting broader responsible AI principles. The taxonomy of logging purposes provided by Foalem et al. (2024) and the proposed ML observability framework from \cite{shankar2021towards} highlight the potential for logging practices to evolve beyond debugging and operational monitoring, toward ensuring ethical compliance and enhancing trust in ML systems.

This study extends existing research by identifying critical gaps in current logging practices for responsible AI. It contributes to the development of tailored logging strategies that systematically capture and audit metrics related to fairness, explainability, and privacy. By leveraging insights from prior work, such as the observability challenges outlined by \cite{shankar2021towards}, our findings aim to bridge the gap between practical logging needs and the overarching goals of responsible AI.

\subsection{Logging in Traditional System}
The study of logging in traditional systems has been widely explored in software engineering, focusing on improving logging practices, optimizing log quality, and addressing common challenges. In this section, we provide a structured overview of related work, categorizing studies into logging practices, guiding logging decisions, and the quality of log statements.

\textbf{Logging Practices, Benefits, and Challenges}: 
Logging has been extensively studied in traditional systems, including web-based systems \citep{chen2017characterizing}, Android applications \citep{zeng2019studying}, and database systems \citep{liu2009framework}. Researchers have highlighted multiple benefits of logging, such as system comprehension, monitoring, failure diagnosis, and auditing \citep{batoun2024literature}. Logs help developers understand system behavior, diagnose issues, and improve system reliability \citep{batoun2024literature, butin2014log}.
Prior studies have identified major challenges in logging, including balancing logging verbosity with system performance, log storage management, and information overload \citep{batoun2024literature}. Additionally, logs frequently contain missing or inconsistent execution data, making debugging and auditing more difficult \citep{batoun2024literature}.

\textbf{Guiding Logging Decisions}:
To address logging challenges, researchers have proposed several techniques to guide logging decisions, particularly in determining what to log, where to log, and which log level to use \citep{zhu2015learning, li2017log}. Machine learning and rule-based approaches have been developed to help developers optimize their logging practices.
For example, automated log placement tools suggest locations within the code where logging is most beneficial. Some studies also focus on recommending log levels to ensure that critical events are appropriately recorded without introducing excessive logging overhead \citep{mastropaolo2022using}. These tools and methodologies help ensure that logs remain relevant and useful throughout the software lifecycle.

\textbf{Quality of Log Statements}:
Several studies have investigated the quality of log statements, identifying logging anti-patterns and log-related vulnerabilities \citep{aghili2024empirical, butin2014log}. Common issues include logging redundant or irrelevant information, improper log formatting, and incorrect log levels \citep{batoun2024literature}. To mitigate these issues, researchers have developed automated tools for log quality assessment, log duplication detection, and static analysis of logging code \citep{chen2017characterizing, li2021studying, li2017log}.

In contrast, our research focuses specifically on logging practices in machine learning-based systems and their alignment with responsible AI principles, exploring how logging can effectively support auditing these principles.

\section{Threats to validity}
\label{sec:threats_to_validity}
In this section, we discuss the potential threats to the validity of our research
methodology and findings.

\textit{Internal Validity.} The subjective judgment of those conducting the manual analyses can pose a threat to the internal validity of our results. To address this concern, manual analyses were conducted by the first three authors of this paper, each with extensive industry and academic experience in ML applications engineering. In cases where there were disagreements among the researchers, a fourth researcher joined in a group discussion until reaching a consensus. We believe that this approach significantly reduced the likelihood of introducing false positives in our analyses. Nonetheless, future replications and extensions of our work are desirable. Our replication package contains all the data and scripts used in our study. It is available online at \citep{replication}.

\textit{Construct Validity.} To ensure the construct validity of our research, we needed to avoid any errors that could arise from extracting logging statements and responsible ML call functions, including those that may have been commented out by developers. Therefore, we developed a static code analyzer on top of the widely used Python AST parser for static analysis of Python code through official documentation\footnote{\url{https://docs.python.org/3/library/ast.html}} \citep{d2016collective, dilhara2021understanding}. This analyzer only extracts uncommented statements, allowing us to avoid collecting logging statements and responsible ML tools call functions that were commented out by developers. Although we selected valid, responsible ML tools using a systematic approach, it is possible that we might have missed some tools. However, by sorting the search results based on popularity, we expect no important tool to be left during the investigation. Additionally, our study emphasizes the role of logging in auditing responsible AI principles, it is important to recognize that logging is not the sole mechanism for achieving this goal. Other approaches, such as testing, runtime monitoring, or external audits, can complement or, in some cases, substitute logging practices  \citep{riccio2020testing, zhang2020machine}. However, logging provides a way for continuous auditing of the principles; as our results in RQ3 show. The results in \autoref{fig:Logging_diagram_cropped} suggest that certain metrics are underrepresented in current logging practices. 
The discrepancy between practitioners' positive feedback on the framework (Table \ref{tab:survey_validation}) and their limited implementation of the metrics (Figure \ref{fig:Logging_diagram_cropped}) points to potential barriers, such as technical constraints, organizational priorities, or lack of awareness, which may hinder broader adoption as mentioned by some practitioners in RQ2 result section. These barriers were not fully explored in this study and present an opportunity for future research to investigate how logging interacts with other methodologies, such as testing and monitoring, to provide a more comprehensive and effective approach to auditing responsible AI applications.

\textit{External Validity.} In this study, we selected a sample of open-source ML-based projects written in Python hosted on GitHub to investigate our research questions. However, the selection of these projects may be threatened, as our findings and results may not be generalized for proprietary systems, and systems written in other programming languages. This way, further studies can be performed exploring different programming languages and gaining insights into different logging requirements for auditing responsible ML-based applications used by ML practitioners. Nevertheless, given that Python is commonly used for ML-based application development~\citet{dilhara2021understanding}, our study provides valuable insights into current practices in logging for auditing ML-based systems.  Additionally, while our study focused on widely recognized responsible AI principles such as Privacy, Fairness, Transparency \& Explainability, and Security \& Safety, it does not explicitly address principles like Accountability and Inclusiveness. This absence was due to the lack of explicit library features addressing these principles in the tools analyzed. Furthermore, while we made a concerted effort to design a comprehensive query for identifying responsible ML libraries, it is possible that some responsible AI principles, particularly company-specific or self-defined principles such as ``Inclusiveness" from Microsoft's Responsible AI guidelines, were not considered in our study. Our search terms were collaboratively selected based on universally recognized principles like fairness, privacy, transparency, explainability, interpretability, and accountability, as these are widely discussed in both research and practical applications. However, the omission of other principles may limit the scope and generalizability of our findings, the selected principles cover the most critical aspects of responsible AI commonly emphasized in both research and industry \citep{de2021companies}.
Future research could expand the list of principles and include company-specific terminologies to achieve a more holistic understanding of responsible AI principles and their implementations.

\section{Conclusion and future work}
\label{sec:conclusion}
In this study, we aimed to shed light on the usage of logging practices, to collect responsible ML-related information in machine learning (ML) systems, thereby enabling continuous auditing of responsible AI principles. Our results revealed that privacy, interpretability \& explainability, fairness, and security and safety are four responsible AI principles that ML practitioners frequently implement in their systems. Additionally, from our analysis in RQ2, we found that while practitioners frequently log performance metrics such as accuracy and loss values, they infrequently document critical metrics related to fairness, explainability, privacy, and security. This discrepancy highlights areas where logging practices can be improved to better support responsible AI initiatives.

Furthermore, certain metrics related to those responsible ML principles, such as group or individual fairness, SHAP values, and privacy metrics, are suggested to be regularly monitored and logged to prevent bias, ensure transparency \& explainability, and maintain privacy in ML applications. Our findings offer valuable insights not only for ML practitioners and logging tool developers but also for researchers looking to improve existing logging practices or develop new ones that support the logging of responsible ML principles and facilitate the auditing of ML applications through logging.

Overall, our study contributes to the ongoing efforts on responsible AI and provides recommendations for enhancing the logging practices of ML applications, to ensure compliance with ethical and legal standards. Future work could include the development of automated tools for logging and auditing machine learning models, and conducting a more extensive qualitative analysis by surveying more practitioners about this research topic.

\section*{ }
\label{sec:declaration}

\noindent\textbf{Data availability statement:} Dataset and source code used in our paper is publicly available online, [\url{https://shorturl.at/FLg9z}].

\section*{Declarations}
\label{sec:declaration}

\textbf{Conflicts of Interest:} The authors declare that they have no conflicts of interest relevant to the content of this article.

\textbf{Ethics Declaration:} This study was approved by Ethics Committee of Polytechnique Montreal (CER-2324-25-D). All participants provided informed consent prior to their participation.

\textbf{Informed Consent:} All participants provided informed consent before taking part in the survey.

\textbf{Author Contributions:} 
\begin{itemize}
  \item \textbf{Patrick Loic Foalem:} Conceptualization, Data Curation, Formal Analysis, Investigation, Methodology, Visualization, Writing -- Original Draft, and Writing -- Review \& Editing.
  \item \textbf{Leuson Da Silva:} Writing -- Review \& Editing.
  \item \textbf{Foutse Khomh:} Project Administration, Resources, Supervision, Validation, and Writing -- Review \& Editing.
  \item \textbf{Heng Li:} Supervision and Writing -- Review \& Editing.
  \item \textbf{Ettore Merlo:} Supervision and Writing -- Review \& Editing.
\end{itemize}



\bibliographystyle{plainnat}
\bibliography{Paper}

\end{document}